\documentclass[12pt]{article}%
\usepackage{amsfonts}
\usepackage{amssymb}
\usepackage{setspace}
\usepackage[round]{natbib}
\usepackage{amsmath}%
\usepackage{amsthm}%
\usepackage{relsize}
\usepackage{subcaption}
\usepackage{xcolor}
\usepackage{booktabs}

\usepackage{microtype}
\usepackage{paralist}

\usepackage{comment}
\usepackage{graphicx}

\usepackage[top=7pc,bottom=7pc,left=8.4pc,right=8.4pc]{geometry}

\usepackage{hyperref}

\usepackage[plain]{algorithm}
\usepackage{algorithmic}

\newtheorem{theorem}{Theorem}

\newtheorem{corollary}[theorem]{Corollary}

\usepackage{mathtools}

\title{Sequential nonparametric tests for a change in distribution: an application to detecting radiological anomalies}
\author{Oscar Hernan Madrid Padilla\footnote{Ph.D student, Department of Statistics and Data Sciences, University of Texas at Austin.  oscar.madrid@utexas.edu} \\
Alex Athey\footnote{Research Scientist, Applied Research Laboratories, University of Texas at Austin, alex.athey@arlut.utexas.edu} \\
Alex Reinhart\footnote{Ph.D student, Department of Statistics, Carnegie Mellon University. areinhar@stat.cmu.edu}\\
James G.~Scott\footnote{Associate Professor, Department of Statistics and Data Sciences and McCombs School of Business, University of Texas at Austin.  james.scott@mccombs.utexas.edu}}
\date{\today}

\begin{document}

\maketitle

\begin{abstract}
We propose a sequential nonparametric test for detecting a change in distribution, based on windowed Kolmogorov--Smirnov statistics.  The approach is simple, robust, highly computationally efficient, easy to calibrate, and requires no parametric assumptions about the underlying null and alternative distributions.  We show that both the false-alarm rate and the power of our procedure are amenable to rigorous analysis, and that the method outperforms existing sequential testing procedures in practice.  We then apply the method to the problem of detecting radiological anomalies, using data collected from measurements of the background gamma-radiation spectrum on a large university campus.  In this context, the proposed method leads to substantial improvements in time-to-detection for the kind of radiological anomalies of interest in law-enforcement and border-security applications.

\bigskip

\vspace*{-.3cm}
\noindent Key words: sequential testing, Kolmogorov--Smirnov test, anomaly detection, radiological survey
\end{abstract}

\newpage

\section{Introduction}

\subsection{Finding radiation anomalies}

Radiologically active materials are used widely in industry, medicine, and research.  Yet an unsecured, lost, or stolen radiological source can present a major threat to public safety \citep{MarkGaffigan,varley2015remediating}.   To deal with the potential environmental and security hazards posed by such a scenario, the United States government uses various detection procedures at ports of entry to the country \citep{ryan2014predicting}.  Moreover, security agencies that try to prevent terrorist attacks are keenly interested in the problem of identifying and locating stolen or smuggled radiation samples \citep{zelakiewicz2011soris,MarkGaffigan,ryan2014predicting,alamaniotis2015anomaly}.  Even at the local level, police departments have shown increasing interest in the deployment of systems for detecting anomalous radiological sources \citep{tansey2015multiscale}.

Statistically speaking, the radiological anomaly-detection problem is one of detecting a change in distribution.  Sequential data is collected from a sensor that measures the energies of arriving gamma rays.  These observed energies $y_{t,i}$ are random variables drawn from an energy spectrum, which is a probability distribution over the set of possible gamma-ray energies.  The question is whether those measured energies are from the normal background spectrum, and therefore harmless, or whether they are from an anomalous spectrum due to the presence of a nearby radiological source.  If the data suggest the presence of an anomaly, an alarm is raised, and some further action is taken.

The key methodological questions are how to measure departures from the normal background gamma spectrum, and how to choose a stopping rule, or equivalently a threshold for declaring an alarm.  Clearly the observer must use a stopping rule that balances the rate of false alarms against the possibility of drastic harm caused by missing a true anomaly.  A key practical challenge is that neither the background spectrum nor the spectra of likely anomalies have any particular parametric form.

In this paper, we study the question of how to construct a stopping rule for detecting a change in distribution based on streaming data, without parametric assumptions.  Our proposed stopping rule is based on a sequential version of the classical Kolmogorov--Smirnov (KS) test, which we apply to radiological survey data collected over several weeks from the campus of the University of Texas at Austin.  The paper emphasizes three key properties of the proposed approach.
\begin{enumerate}
\item Our test is both extremely simple in practice and, as we show, amenable to rigorous analysis of both its false-alarm rate and its power.
\item The stopping rule for our method depends only on an easily understood, user-defined tolerance for false alarms, and not on any underlying properties of the data-generating process.  (See Corollary \ref{cor1}.)  In that sense, our procedure is analogous to a traditional (non-sequential) test based on a pivotal quantity.  This is not true of other all other existing methods, some of which involve test statistics whose behavior depends on the null.
\item Most importantly, the method leads to large practical improvements over existing methods for radiological anomaly detection.  For hard-to-detect anomalies whose spectra are poorly separated from the background---see the cobalt example in Table \ref{realisticSimulation}---our test improves time to detection by nearly an order of magnitude over existing methods.
\end{enumerate}
 

\subsection{Our data}
 
 Our primary data source consists of measurements of the background gamma-ray spectrum that were collected during July and August 2012 at the University of Texas J.J. Pickle Research Campus (hereafter, PRC).  The measurements were taken with a cesium-iodide scintillator detector hooked up to a GPS unit and driven around campus in a golf cart.  This enabled us to characterize the background radiation over a wide spatial area.  Our secondary data source consists of a small field experiment, described in Section 5.
 
Our detector records the observed gamma ray energies in each two-second window, and discretizes these energies into an extremely fine histogram comprising 4,096 discrete bins, where higher-numbed bins correspond to higher energies.  We winsorize the data at bin 2,048:  most counts in the highest 2,048 bins are overwhelmingly due to astronomical sources of high-energy gamma radiation, meaning that fine-scale structure in this part of the spectrum is not very helpful for detecting a terrestrial anomaly.   Despite this very fine discretization, there is effectively no practical difference from a truly continuous setting.  See Section \ref{sec:properties_of_ks} for a more detailed discussion of this issue.

\begin{figure}
 		\begin{center}
 			\includegraphics[width=6in,height= 3in]{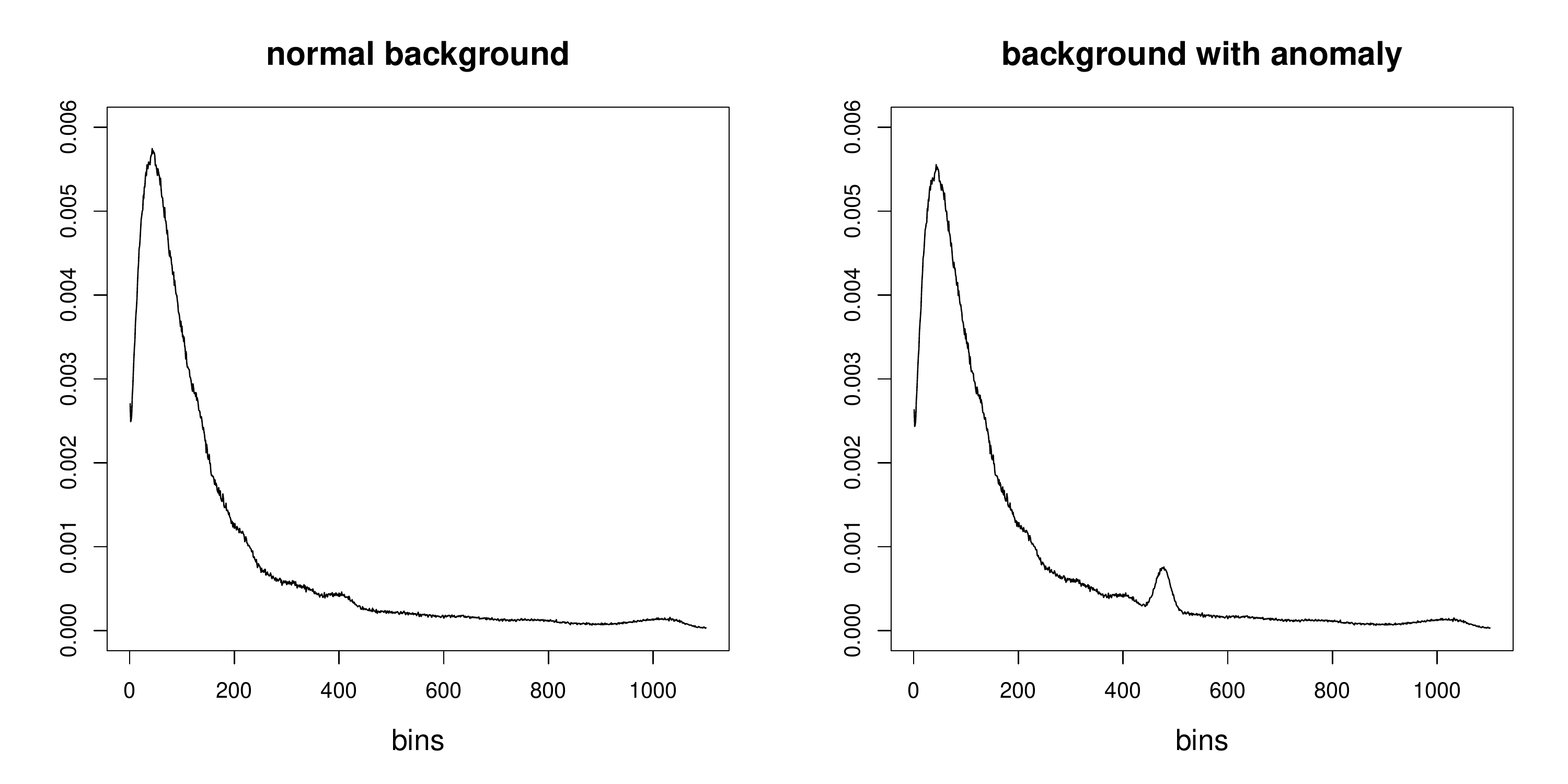}
 				\includegraphics[width=6in,height= 3in]{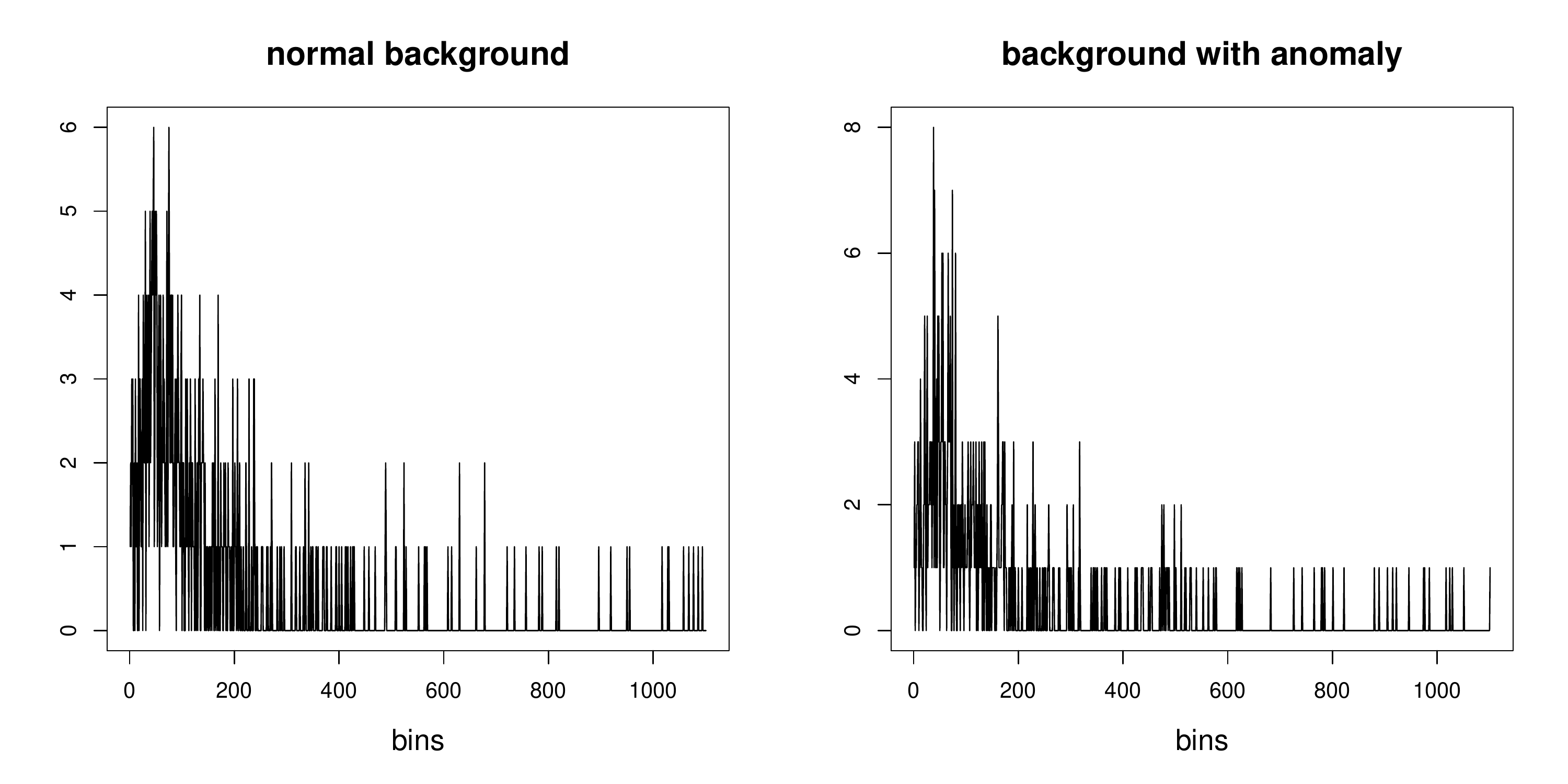}  
                        \caption{\label{example} The top left panel shows an example density corresponding to a normal background radiation at the PRC, while the top right panel shows the same background but with the presence of an anomaly: a 100 milliCurie source of Cesium 137 isotope located at distance of 150m from the detector.   The bottom two panels show histograms of 500 observations from the densities in the top two panels. These represent an example of the raw data collected sequentially with the detector, and illustrate the difficulty of the radiological anomaly detection problem.  }
 		\end{center}
 	\end{figure}
 	
 Figure \ref{example} gives the reader some idea of the challenge posed by the anomaly-detection problem, in the context of our data, by showing an example of a normal regime and one in which there is an anomaly present. The first panel in Figure \ref{example} shows the background energy spectrum at a specific location at the PRC, as a black line.  This has been estimated using the multiscale technique of \citet{tansey2015multiscale}.  The second panel shows a situation where there is an anomalous radiation source mixing with the normal regime.  Specifically, this is the theoretical spectrum an observer would measure with our detector in the presence of a 100 milliCurie Cesium-137 anomaly located at a distance of 150 meters; this anomaly accounts for the bump near bin 550.  However, the bump is much less apparent in the noisy data of the bottom two panels.  At this distance, the anomalous source is difficult to distinguish from the background on the basis of even 500 observations.  This noise is characteristic of the data collected in the field under operational conditions and realistic time constraints.

\subsection{Outline} 

Our goal is to construct a protocol that can detect a change in distribution as quickly as possible while controlling the number of false alarms. To that end, we construct a sequential anomaly detection procedure based on the classical Kolmogorov--Smirnov test for equality of continuous distributions. We establish the statistical properties of our stopping rule, such as a lower bound for the power of the rule when detecting unknown anomalous radiation levels. The proposed approach is highly flexible, in that it can directly work with the raw observations $\{y_{t,i}\}$, where available, or with the binned counts from the spectrometer.  We then show that the method outperforms existing approaches both on simulated and real data.
	
The rest of the paper is organized as follows. Section 2 starts by putting our sequential anomaly detection problem into a statistical framework and discussing its relation with previous work in the literature. In Section 3.1 we revisit the classical KS test and its discrete variants. Section 3.2 then briefly summarizes the previously known  sequential KS procedure. This is leads to the exposition of our method in Section 3.3, presenting two views of the test that allow to handle raw photon measurements and binned counts. We then move to study the statistical properties of our approach in Section 3.4.  Section 4 benchmarks our proposed method on a variety of simulated examples.  Section 5 contains our main application to radiological anomaly detection, using the data described in the introduction.  Finally, Section 6 summarizes our contributions and discusses some limitations of the approach.

\section{Background on radiological anomaly detection}

\subsection{The assumed model}

Although photons arrive in continuous time, a typical radiation detector outputs measurements in discrete time.  While our approach can handle continuous-time measurements with no additional complication, hereafter we assume that time is discrete, and that possibly multiple photons arrive at each time point.

We let $\{y_{t,i} \}_{i=1}^{N_t}$ denote the set of measured energies from the gamma rays arriving at time $t$.  The underlying physics imply that
\begin{equation}
\label{sampling_model}
y_{t,i}  \stackrel{\text{iid}}{\sim}  f_t \, , \; i = 1,\ldots, N_{t} \, , \quad N_{t} \sim \text{Poisson}( \mu ) \, , 
\end{equation}
for some overall count rate $\mu > 0$, where $f_t$  is the gamma-ray spectrum at time $t$, i.e.~a probability distribution with sample space $\Omega \subset \mathbb{R}^{+}$.

One simple approach to detect anomalies is to measure overall radiation levels, i.e.~to compare the overall count $N_t$ with the background rate $\mu$.  But a major problem with this approach is that different devices have different intrinsic sensitivities to radiation, and a search for an anomaly might involve a different device than was used to measure the background.   Moreover, in our background data, we have observed noticeable differences in overall counts observed using the \textit{same} detector from one day to the next.  This may be caused by changes in the detector orientation or different heights of the detector above the ground (depending on who was carrying it).

These facts imply that $\mu$ is not merely a property of the background, but a property of the background and the measurement device/process considered jointly.  Thus from an operational standpoint, attempting to detect anomalies using $N_t$ is fraught with difficulties.   It is much more robust and effective to monitor the energies $y_{t,i}$ of the detected photons, which (after laboratory calibration) are independent of the device and more sensitive to anomalous sources.

Therefore we frame the radiological anomaly-detection problem in terms of a change in distribution: at some unknown time $v$ during the search process, there is a change in the probability distribution of measured photon energies. Explicitly, there exist two pdf's $f_0$ and $f_c$ with sample space $\Omega$ such that 
\[
f_t \,\,=\,\,\begin{cases}
f_0   & \text{for}\,\,  t \leq  v\\
f_c   & \text{for}\,\,  t > v.\\
\end{cases}
\]
We refer to $f_0$ as the pre-change density, which is known; $f_c$  as the post-change density, which is unknown; and $v$ as the changepoint, also unknown.  The classic approach to detecting a change in distribution, due to \cite{pollak1987average}, is based on a likelihood-ratio statistic, and requires the assumption that $f_c$ is in an exponential family.  But in our case, neither $f_0$ nor $f_c$ take any particular parametric form, necessitating a different approach.

 In our application, the changepoint $v$ is the moment when the observer starts to measure photons from an anomalous source---for example, as a result of driving and suddenly passing by an area containing an illicit radiation source.  When this happens, the radiation emanating from the unknown anomalous source mixes with the background radiation.   This implies that the post-change density satisfies   
 \begin{equation}
 \label{f_c}
 f_c \,\, = \,\,  w\,f_0  \,+ \, (1-w)\,f_A, 
 \end{equation}
 where $w \in (0,1)$, and $f_A$ represents a pdf associated with photon measurements from a ``pure'' anomalous source.
 
 Both $w$ and $f_A$ are unknown. The weight $w$ is a function of many variables, including the distance from the detector to the anomalous source and the size or intensity of the source.  The density $f_A$  depends on the emission spectrum of the source, together with the physics that govern the scattering of photons as they interact with the surrounding environment on their way from source to observer.  Although the emission spectra of likely anomalies are known precisely, this does not help in practice: $f_A$ does not refer to the emission spectrum of the anomaly, but rather what an observer would actually see from that anomaly in the absence of background.  In a controlled laboratory environment, the contribution of an anomalous radiation source to the measured spectrum has a simple and known dependency  based on distance from the observer and attenuation in air.   However, in a complex urban environment, the range of densities of different building materials, together the complex geometry of structures in the built environment, combine to produce an extremely complex, unknown transfer function between the source emission and the detected emission at any given location.  Thus while (\ref{f_c}) helps us to understand the problem, it does not explicitly help to detect anomalies, and is not a formal assumption of our method.

\subsection{Connections with previous work}

In recent years, radiological anomaly detection has attracted significant attention from the research community.  Broadly speaking, there are two typical anomaly-detection scenarios: (1) a stationary detector, such as those at ports or border crossings; and (2) mobile detectors, of the kind used by law enforcement to screen for illicit radiological sources (like dirty bombs).  The method we propose is appropriate for either scenario, although the real-data examples we consider in this paper all involve mobile detectors.

\paragraph{Characterizing the background spectrum.} 

In the mobile detector scenario, law-enforcement personnel visit different sites within a search area and take radiation measurements.  These (noisy) measured spectra can then be compared against the background spectrum $f_0$, which varies spatially due to differences in naturally occurring radioactive material (NORM) in the natural and built environments.  In much of the radiation literature, the background is determined from recent observations as the detector moves.  These measurements are often smoothed using a moving average or Kalman filter \citep[e.g.][]{pfund2006examination}.

However, much of the recent work in this area focuses on using previous background observations to improve detection sensitivity \citep{vetter:etal:2015}.  In this type of system, the detection phase is preceded by a long-term mapping phase---for example, the kind of radiological survey described in \citet{reinhart2014spatially}. This means that each measurement can be compared with the background at the current location, rather than some global average background or the average spectrum of the last ten minutes.  For many other references on this idea, see \citet{tansey2015multiscale}.  Throughout the paper, we assume that the spatial variation background has been fully characterized using historical radiological survey data (which is, in fact, available for our real-data examples), and we therefore use the background at the current location for $f_0$.

\paragraph{Detecting anomalies.}  We highlight two different kinds of approaches to radiological anomaly detection: retrospective and sequential.

Retrospective methods analyze a batch of data collected in the past.  Both \cite{chan2014distribution} and \cite{reinhart2015detecting} independently established a connection between retrospective radiation detection and the classical Kolmogorov--Smirnov test, which was also used by \cite{tansey2015multiscale}.
Several other retrospective approaches have been built upon the  Spectral Comparison Ratio (SCR) statistic, introduced by \cite{ pfund2006examination}. This includes work by \cite{du2010noise} and \citet{reinhart2014spatially}.   These methods compare a transformed test-count vector with a known background vector, and assume that the null distribution of this test vector is approximately normal.  This implies a chi-square distribution on a dissimilarity measure between observations and background.   There have also been efforts to apply this framework to the sequential anomaly detection problem.   For example, \cite{pfund2010low} proposed an extension of the SCR algorithm to the sequential framework. However, it is not entirely clear how to select some of the method's tuning parameters. Moreover, our experiments show that the method is underpowered compared to the proposal considered here, particularly if the anomaly is poorly separated from the background, or if the anomalous source is located far from the detector.


Finally, there are other possible anomaly detection methods available that exploit the discreteness of the measurement process.  Recall that our photon measurements $\{y_{t,i} \}_{i=1}^{N_t}$ are binned into a very fine histogram at each time $t$.  This produces a vector of counts $x_t = (x_{t,1}, \ldots, x_{t,D})$, where $D$ is typically large (in our case, 2048).  If we assume without loss of generality that the sample space $\Omega = [0,1]$, we can partition $\Omega$ into discrete energy channels $\{B_j\}_{j=1}^D$ as
\begin{equation}
\label{partition_definition}
B_j \,\,=\,\, [(j-1)/D,j/D),\,\,\,\,\text{for }\,\,\,j = 1,\ldots, D-1, \,\,\,\,\text{and}\,\,\,B_D = [(D-1)/D, 1].
\end{equation}
The count in energy channel $j$ is
\[
x_{t,j} = \left\vert \left\{   i \,\,:\,\,  y_{t,i} \in B_j ,\,\,1 \leq i \leq N_t   \right\}\right\vert,\,\,\forall j \in \{1,\ldots, D\}.
\]

Given the sampling model in (\ref{sampling_model}), one immediate avenue to study the changepoint detection problem is to assume that 
\begin{equation}
\label{model_before_change}
x_{t,j} \sim \mbox{Poisson}(\lambda^{(0)}_{j}) \, ,
\end{equation}
i.e.~the events in channel $j$ happen at baseline rate $\lambda^{(0)}_j$.  Here $\lambda^{(0)}$ is known beforehand.  After the unknown change point, i.e.~when $t > v$, the counts are generated independently as
\begin{equation}
\label{model_after_change}
x_{t,j} \sim \mbox{Poisson}(\lambda^{(c)}_{j}) \, ,
\end{equation}
where the vector $\lambda^{(c)}$ is unknown.   From (\ref{model_before_change}) and (\ref{model_after_change}), it would seem that, in principle, any sequential changepoint-detection procedure that applies to a Poisson distribution, and that can handle an unknown post-change rate, could  be used to construct a radiological anomaly-detection procedure.

The classic paper on detecting a change in distribution for exponential families is by \citet{pollak1987average}, who proposes to compute, at every time, a retrospective log-likelihood ratio. \citep[We refer to this method as EF; see also][for a generalization]{mei2006sequential}. Unfortunately, the theoretical guarantees in this paper are limited to the one-dimensional case, and to non-lattice distributions. Both explicitly exclude the model in (\ref{model_before_change}), which involves a multivariate vector of Poisson observations.  A version of the test in \citet{pollak1987average} can still be applied, but its formal properties have not, to our knowledge, been characterized yet.  Methods based on generalized likelihood ratios (GLR) have been studied by several authors \citep{siegmund1995using,basseville1993detection,lai1995sequential}.

We will use these sequential methods based on the Poisson assumption (\ref{model_before_change}--\ref{model_after_change}) as benchmarks to study the power of the proposed approach.  However, we point out that they all suffer from the same practical difficulty as a test based on a comparison of $N_t$ with $\mu$, the overall radiation level.  Specifically, they require that the detector used to search for anomalies has the same overall sensitivity to radiation as the device used to measure the background, so that the per-channel Poisson rates are comparable.  In most practical search scenarios, this cannot be guaranteed.  This is an important practical reason to prefer the KS-based approach proposed here.

\section{Stopping rules from sequential KS tests}

\label{kolmogrov_test_section}

We now construct stopping rules for the anomaly-detection problem described in the introduction. Our method builds upon the classical Kolmogorov--Smirnov (KS) test \citep[e.g.][]{massey1951kolmogorov}. This is a popular non-parametric method used  to assess whether a specific sample has been drawn from a given continuous distribution.  We note that, in the context of radiological anomaly detection, \cite{chan2014distribution} and \cite{reinhart2015detecting} have successfully applied \emph{retrospective} KS tests for the radiological anomaly-detection problem.  Our goal in this paper is to construct a \emph{sequential} version of this test, and to place the choice of stopping rule for this test on a firm statistical foundation.  In that sense, we provide a nonparametric counterpart to the parametric sequential method of \citet{pollak1987average}.

\subsection{The classical KS statistic and a discrete variant}

To provide some context for our proposed approach, we first describe an instantaneous or single-window approach to anomaly detection based on the KS statistic.  Given the sampling model in (\ref{sampling_model}), we denote by $F_0$ the cumulative distribution function (CDF)  associated with $f_0$. We also write  $\hat{F}_t$ for the empirical CDF derived from the observations $\{y_{t,i}\}_{i=1}^{N_t}$ from time window $t$:
\[
   \hat{F}_{t}(y) \,\,=\,\,  \frac{1}{N_t}\,\sum_{i=1}^{N_t}\, \bold{1}_{ (-\infty, y_{t,i} ] }(y),
\]
for all $y \in \mathbb{R}$, where $\bold{1}_{A} $ is the set indicator function of $A$.    It is well known that, for large $N_t$, the one-sample KS statistic
\begin{equation}
\label{Dt}
   D_t \,\,\,= \,\,\,\sqrt{N_t}\, \,\sup_y \, \left\vert F_0(y)  - \hat{F}_t(y) \right\vert,
\end{equation}
converges in distribution to the Kolmogorov distribution.

For the binned-counts model from (\ref{model_before_change}) and (\ref{model_after_change}), an analogous statistic can be constructed.   We begin with the normalized vector of rates 
\begin{align*}
w^{(0)} & =   \left( \frac{\lambda_1^{(0)} }{\lambda^{(0)}_{\cdot}}, \ldots, \frac{\lambda_D^{(0)} }{\lambda^{(0)}_{\cdot}}    \right) \\
w^{(1)} & =   \left( \frac{\lambda_1^{(c)} }{\lambda^{(c)}_{\cdot}}, \ldots, \frac{\lambda_D^{(c)} }{\lambda^{(c)}_{\cdot}}    \right)
\end{align*}
with
\[
    \lambda^{(0)}_{\cdot}   \,\,=\,\, \sum_{j=1}^{D} \lambda^{(0)}_{j},\,\,\,\,\,\,\,\,   \lambda^{(c)}_{\cdot}   \,\,=\,\, \sum_{j=1}^{D} \lambda^{(c)}_{j}. 
\]
From our Poisson  model, Equations (\ref{model_before_change}) and (\ref{model_after_change}), we have
$$
(x_t \mid N_t) \sim \mbox{Multinom}(N_t, w^{(1)}) \, , \quad w^{(1)} \neq w^{(0)} \,,
$$
where $N_t =  \sum_{j=1}^D x_{t,j}$, and where the new probability vector $w^{(1)}$ is unknown.

We now let
$$
F^{0}(j) = \sum_{k=1}^j w_{k}^{(0)},
$$
denote the cumulative sum of the probabilities in the first $j$ channels.  This is like a discrete approximation to $F_0$.  Similarly, define
$$
\hat{F}^t(j) = \frac{1}{N_t}\sum_{k=1}^j x_{t,k},
$$
as the cumulative frequency of the counts at time $t$ across the first $j$ energy channels.  Next, define a discrete version of the KS statistic as
\begin{equation}
\label{KS_test_statistic}
        \Delta_t = \sqrt{N_t} \cdot \,\max_{1 \leq j \leq D} | F^{0}(j) - \hat{F}^t(j) |,
\end{equation}
as the (rescaled) maximum of the absolute differences between the background CDF and the empirical CDF at time $t$.  Under the normal regime, $\Delta_t$ will very nearly have a Kolmogorov distribution, whereas it will have a different (though unknown) distribution if there is a change point.

We say ``very nearly'' here because, unlike the statistic $D_t$ defined in (\ref{Dt}), the discretization to bins in (\ref{KS_test_statistic}) might introduce some small departure from the theoretical Kolmogorov distribution, which assumes no ties among the data.  But this bias is virtually undetectable with a large number of bins (e.g.~$>1000$ in our case), and in all of our data sets, the theoretical KS null distribution and the actual null distribution of $\Delta_t$ are identical for all practical purposes.

The statistic $\Delta_t$ was used in \cite{chan2014distribution} and \citet{reinhart2015detecting}. Neither of these approaches, however, yields a satisfactory protocol for sequential real-time analysis of radiological data---only retrospective detection using fixed batches of data.   One truly sequential approach using the KS statistic would be to calculate a $p$-value from $\Delta_t$, and to use any of several recent procedures for controlling the false discovery rate in a sequential-testing framework \citep[e.g.][]{foster2008alpha,javanmard2015online}.   But one obvious drawback of this approach is that the $\Delta_t$ statistic does not pool information across temporally adjacent tests, and therefore has needlessly low detection power.

\subsection{\label{ks_precusor} A predecessor sequential test}

Our work is closest in spirit to that of \cite{hawkins1988retrospective}, who proposed a sequential ``pooled'' KS-based test in the setting of sequentially arriving, one-dimensional observations. In our context of multiple observations per time point, this approach involves computing at every time $t$ the test statistic 
$$\kappa_t  \,\, = \,\,\,\sum_{k=1}^t N_k\, \,\sup_y \, \left\vert F_0(y)  - \hat{F}_{1:t}(y) \right\vert,  $$
where $\hat{F}_{1:t}$ is the empirical CDF between periods time $1$ and $t$:
\[
\hat{F}_{1:t}(y) \,\,=\,\, \frac{1}{ \sum_{k=1}^t N_k}\,\sum_{k=1}^t\sum_{i=1}^{N_k}\, \bold{1}_{ (-\infty, y_{k,i} ] }(y).
\]
Thus we see that data are pooled at each time step since the beginning ($t=1$).  A crucial point is that $\kappa_t$ is not the classical KS test statistic, and does not have a Kolmogorov distribution under the null. This point was made in \cite{hawkins1988retrospective}, who compensated for this fact by choosing a large threshold for rejection.  The resulting stopping rule for raising an alarm is given by
\begin{equation}
\label{precurssor_ks}
	\inf \{ t\,\,:\,\,  \kappa_t > c \},
\end{equation}
where $c >0$. \cite{hawkins1988retrospective} studied the asymptotic behavior of as the threshold   goes to infinity.   In practice, one must select $c$  to control the number of false alarms over a given time period using historical data (or simulated data). 

Unfortunately, we have found that the rule  (\ref{precurssor_ks})  suffers from having limited power. To understand why, first assume that the change point  happens long after the beginning of the data collection ($v \gg1$). Then immediately after the change point, the computation of $\kappa_t$  will be dominated by samples from the normal regime. As a result, (\ref{precurssor_ks}) yields long delays when the pre- and post- change densities are very similar, as it is the case considered in this paper (c.f.~Figure \ref{example}).

To address this concern, the next section presents a different test that will prove to be much more effective.

\subsection{\label{proposed_ks} The proposed test}

\begin{figure}[t!]
 		\begin{center}
 			\includegraphics[width=6in,height= 3.5in]{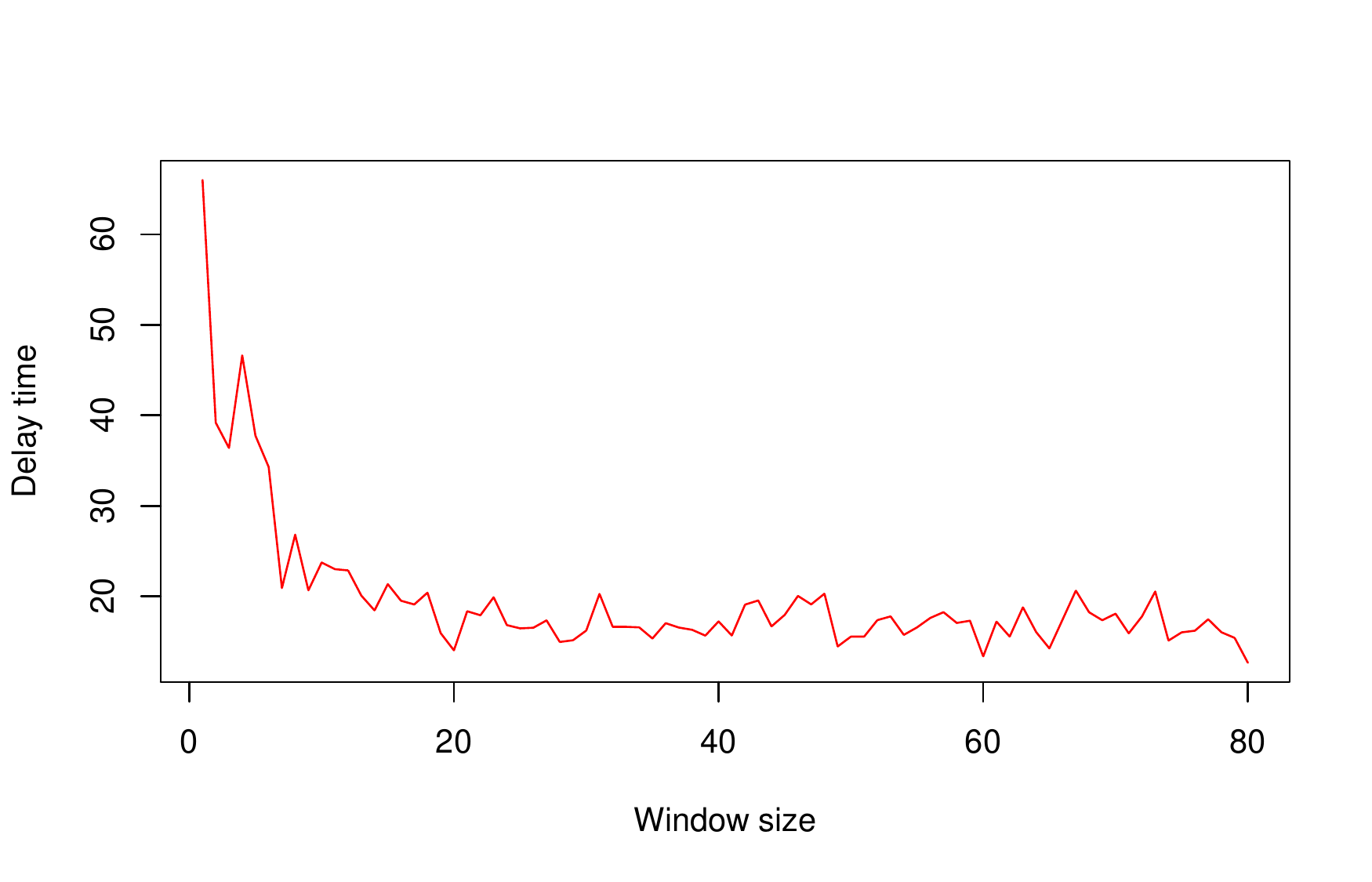} 
                        \caption{\label{ks_power} Delay time for detecting a change point when the pre- and post- change densities are given as in Figure \ref{example}. Here, $E(N_t) = 500$ and we evaluate the performance of our KS method setting the threshold to a value  that would produce an average of one false alarm in an interval of size 1000, estimated on a grid of values using 50 MC simulations. The y-axis shows the delay time in such conditions and the x- axis varies the window size for our Ks method. A window size equals one corresponds to the procedure from \cite{reinhart2015detecting}.   }
 		\end{center}
 	\end{figure}
	
\begin{figure}[t!]
	\begin{center}
		\includegraphics[width=6in,height= 3in]{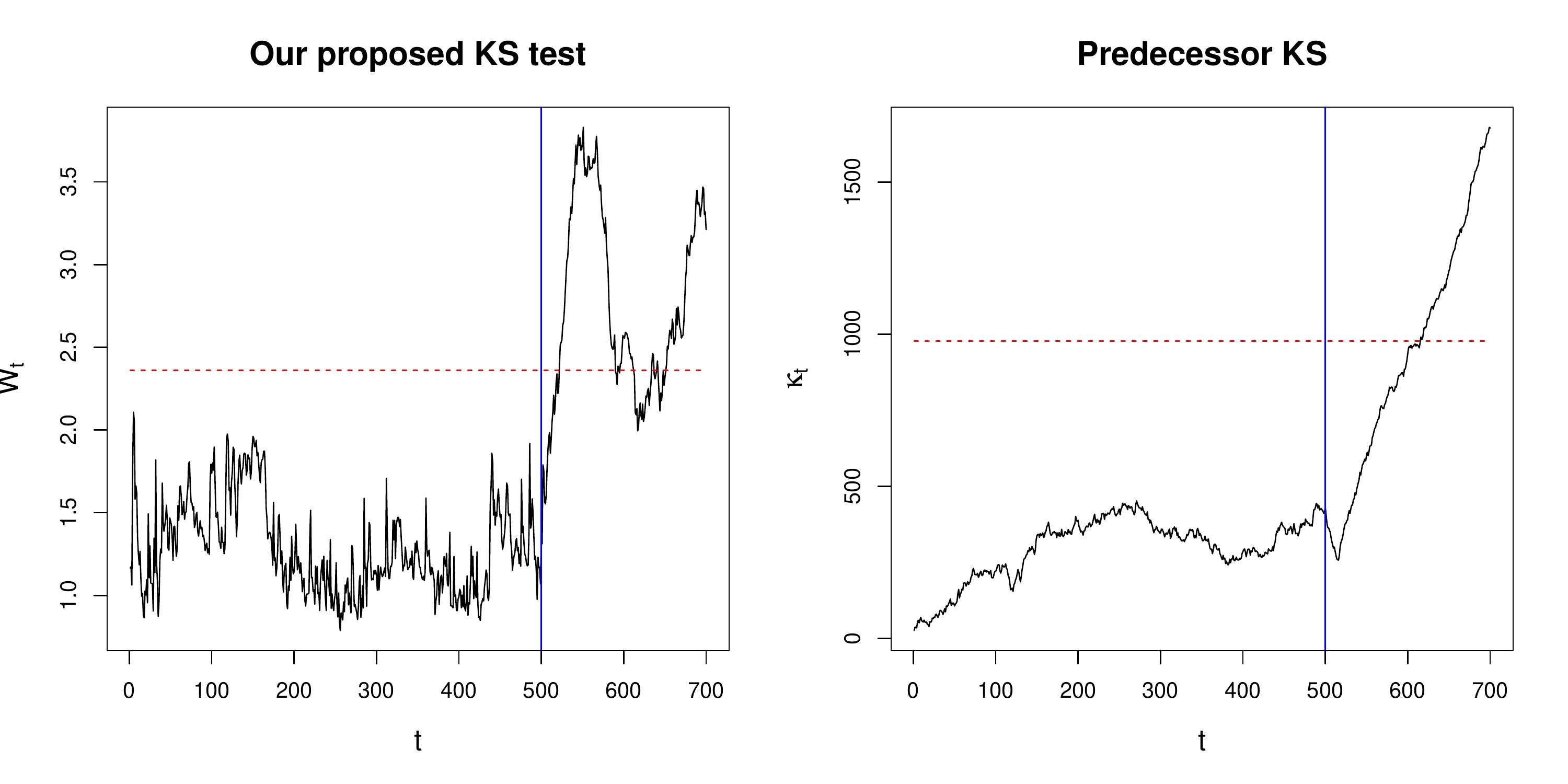} 
                \caption{\label{ks_comparison} Comparisons of the two KS methods proposed in Sections \ref{ks_precusor} and \ref{proposed_ks} when they have been calibrated to have one false alarm in an interval of length 1000.  The first and second panels show the statistics $W_t$  and $\kappa_t$ as a function of time respectively. In each panel the vertical line represents a change point occurring at time 500. The horizontal line is the false alarm threshold. For the predecessor pooled KS the threshold has been chosen, using 50 MC simulations, to a value that leads to an expected value of one false alarm in an interval of length 1000. For our KS method the threshold is chosen according to Corollary \ref{cor1} in the next section such that the expected number of false alarms in an interval of length $1000$ is less than or equal to 1.   The data before and after  the change point is drawn from the pre-- and post--change densities illustrated in Figure \ref{example}. Also, at every time $t$ the number of photon measurements is Poisson distributed with mean $500$. }
	\end{center}
\end{figure}

The idea of our proposed approach is simple: pool data across an increasingly large series of backward-looking windows, and consider the maximal KS statistic over those windows.  To that end, for $1 \leq s \leq t$ and $1 \leq j \leq D$,  we define
\[
   \hat{w}_{s:t,j}  =  \frac{\sum_{k=s}^t x_{k,j} }{\sum_{k = s}^t N_k}.
\]
Next, we construct an empirical CDF  based on samples from time $s$  to time $t$  as
\[
  \hat{F}^{s:t}(j) = \sum_{k=1}^{j} \hat{w}_{s:t,k},
\]
for $j = 1,\ldots,D$.  We then use this to construct a variant of $\Delta_t$ that accounts for all the data collected between time $s$ and $t$:
\[
\Delta_{s:t} = \sqrt{ \sum_{k=s}^{t} N_k }\, \max_{1 \leq  j \leq D} \vert F^0(j) - \hat{F}^{s:t}(j) \vert.
\]
Using this statistic,  we define the window statistic $W$ as
\[
   W_t=  \max_{s: \max \left\{t-L,1\right\}\leq s\leq t} \Delta_{s,t} \, ,
\]
and we propose to stop the data collection process at time
\begin{equation}
\label{kolmogorov_statistic}
\tau_{L} =  \min\left\{t :  W_t  \geq c_{L} \right\},
\end{equation}
where $L \in \mathbb{N}$ and $c_L >0$ are constants. 

The motivation behind (\ref{kolmogorov_statistic}) is that for any given time $t$, we look back and see if there is evidence for a changepoint between times $\max\left\{t-L,1\right\}$ and  $t$. We incorporate the window size $L$ only in order to  have a bound on the overall  complexity for computing the statistic at time $t$.  With this bound, the method has  $O(L\,D)$ cost at each time step.  The idea of choosing a window size to look back and use past data in sequential change-point detection has certainly appeared before in the literature.  However, existing work on this idea has been done in the context of generalized likelihood-ratio tests, and requires parametric assumptions \citep{lai1995sequential}.  This constrains the computational cost of the GLR test, which would go to infinity with its original definition.  While similar in spirit to our proposed approach, this does not involve aggregating counts, nor is it a fully nonparametric test, both of which are practical requirements of our application.

As a special case, $L = 1$ corresponds to a single-window version of the test from \cite{reinhart2015detecting}.   However,  it is much better to choose $L> 1$.  This is empirically illustrated in Figure \ref{ks_power}, where we can see a tremendous improvement in detection time realized by using large values of $L$. A more formal statement of this fact will be given shortly.

Finally, we illustrate with an example that our proposed KS sequential test outperforms the variant of the test from \cite{hawkins1988retrospective} discussed earlier. This is shown in Figure \ref{ks_comparison}, where we see the tremendous gain in power given by our procedure.  The two tests are calibrated to have the same false-alarm rate.  For the pooled KS test from \citet{hawkins1988retrospective}, the threshold has been chosen by Monte Carlo simulation to yield an expected value of one false alarm in an interval of length 1000. For our KS method the threshold is chosen according to Corollary \ref{cor1}, presented in the next section, such that the expected number of false alarms in an interval of length $1000$ is less than or equal to 1.

In the later sections we will provide more comprehensive comparisons.  We now present some important properties of our method. 

\subsection{\label{sec:properties_of_ks}  Properties of the proposed test}

We now turn to the question of how to choose the window size $L$ and the detection threshold $c_L$.  These choices are best illuminated through a careful study of the probabilistic properties of $\Delta_{s:t}$.  To this end, we introduce some notation. Let $F_c$  be  the CDF induced by the post-change density $f_c$, and let
	\[
        d(F_c ,F_0) := \max_{1 \leq j \leq D} \left\vert F_0(\xi_j) - F_c(\xi_j) \right\vert,
	\]
denote the discrete analog of the total-variation distance between the pre-and post-change distributions, where $\xi_j$  is the right end point of  bin $B_j$.  As before, $v$ is the changepoint.

We are now ready to state our first result, which provides an exponential concentration bound on the behavior of $\Delta_{s:t}$. The result makes no assumptions on the pre- and post-change densities, and it only involves the distance $d(F_c,F_0)$. The proof relies heavily on Corollary 1 from \cite{massart1990tight}.

\begin{theorem}
\label{theorem1}
Assume that $s> v$ is fixed. Then

\[
 \lim_{t \rightarrow \infty} \Delta_{s:t}  = \infty \,\,\,\, \text{a.s}.
\]
provided that $d(F_0,F_c) > 0$.  Moreover, for $c_L>0$ the following inequality holds:
\begin{equation}
\label{eqn:ks_concentration}
  \text{P}\left( \Delta_{s:t} > -c_L  + d(F_c,F_0)\sqrt{ \sum_{k=s}^{t} N_k }    \right) \geq 1 - 2\exp(-2\,c_L^2) \, .
\end{equation}
\end{theorem}

As pointed out earlier, for $t\leq v$  the statistic $\Delta_t$ is expected to behave like a Kolmogorov-distributed random variable under the null hypothesis.  This is also the case for $\Delta_{s:t}$ when $s \leq  t \leq v$.  The first part of Theorem  \ref{theorem1} says that, as long as we collect enough data from the abnormal regime, it is possible to (eventually) detect this departure from the null at an arbitrarily high threshold.  The inequality (\ref{eqn:ks_concentration}) sharpens this claim by showing us how quickly the probability of detection approaches 1, as a function of three quantities: the threshold $c_L$, the magnitude of the anomaly (as measured by total-variation distance), and the total number of observations collected over the window from $s$ to $t$.

The following corollary provides a bound of the number of false alarms arising from our test and re-expresses the above inequality in a more immediately useful form. To that end we define, for $T < v$,  
\[
  A_T =  \left\vert \left\{t:  t \leq T,\,\,   \max_{\max\left\{1,t-L\right\}\leq s\leq t} \Delta_{s,t} \geq c_L \right \} \right \vert.
\]
Here $A_T$ can be thought as the number of times that the process would be stopped within a window of length $T$ when there is no change point in $\{1,\ldots,T\}$. 

\begin{corollary}
  \label{cor1}
  If $T \leq v$, then
  \[
  E(A_T) \leq 2\,T\,L\,\exp\left(-2\,c_L^2\right).
  \]
  In addition, if $v < t < v + L$, then
  \[
      P\left(\tau_{L} < t \mid \tau_{L} > v  \right )  \geq 1 - 2\exp(-2\,c_L^2) -  \text{P}\left( \sqrt{\sum_{s=v+1}^t N_s } < \frac{2\,c_L}{d(F_c,F_0)}  \right).
  \]
where $\tau_L$ is given as in Equation (\ref{kolmogorov_statistic}).

\end{corollary}

Corollary $\ref{cor1}$ puts an upper bound on the expected number of false alarms up to time $T$, provided that the change point happens after $T$. This corollary is immediately practical: it can be used to set up the threshold $c_L$ and the window size $L$. For a desirable $T$ and a given tolerance, one can choose the $c_L$  and $L$ so that the expect number of false alarms is less than the tolerance.

We illustrate this with an example calculation of the threshold $c_L$. Assuming that $L= 50$ and $T= 1000$, as in Figure \ref{ks_comparison}, we can set $c_L$ so that the expected number of false alarms to be less than or equal to $1$ over $[1,1000]$. To see this, we invoke Corollary \ref{cor1}:
\[
  \,E(A_T) \,\leq \, 2\,T\,L\,\exp\left(-2\,c_L^2\right)   \, \leq  \, 1, \quad \mbox{for $T=1000$, $L=50$} \, .
\]
This holds as long as 
\[
       c_L \,\geq\, \sqrt{\frac{\log(2\,T\,L)}{2}} \,\,\approx 2.4.
\]
Hence for the example disused in earlier in  Figure \ref{ks_comparison} we set $c_L$  to be $2.4$.

Of course, different combinations of $c_L$ and $L$ can lead to the same tolerance. To have a more informed choice, we can look at the second part of Corollary \ref{cor1}. This says that for a given window  $\delta_t$, the amount of time after the change point $v$ that one is willing to wait to detect the change point, we could constrain $L \geq  \delta_t$. Then for any $\epsilon > 0$, if 
\[
 \frac{\sqrt{\sum_{s=v+1}^{v +\delta_t} N_s } }{2\,c_L} > \frac{1}{\epsilon}  
\] 
holds with high probability within the window $\delta_t$,  then we would be likely to detect any changes that satisfy $d(F_0,F_c)  > \epsilon$.

Finally, we observe that it is possible to define a version of $\Delta_{s,t}$ that directly works with the raw observations $\{y_{t,i}\}$. This is done by  defining the empirical CDF
\[
\hat{F}_{s:t}(y) \,\,=\,\,  \frac{1}{ \sum_{k=s}^{t}  N_{k} }\,\sum_{k=s}^{t}\sum_{i=1}^{N_k}\, \bold{1}_{ (-\infty, y_{k,i} ] }(y).
\]
Then we can set
\begin{equation}
\label{continuous_ks}
D_{s:t} = \sqrt{ \sum_{k=s}^{t} N_k }\,\sup_{ y \in \mathbb{R}} \vert F_0(j) - \hat{F}_{s:t}(j) \vert.
\end{equation}
Hence replacing $\Delta_{s:t}$ by $D_{s:t}$ in (\ref{kolmogorov_statistic}), we obtain a stopping rule that can use the raw observations rather than the discretization. From there it is possible to have results as in Theorem \ref{theorem1} and Corollary \ref{cor1}, replacing $\Delta_{s:t}$  by $D_{s:t}$, and $d(F_0,F_c)$  by the total variation distance between $F_0$  and $F_c$. 


\section{Experiments on simulated data}

We now undertake a comprehensive benchmarking study of our proposed approach, using simulated data.  These simulations have two goals.  First, we will show that our method yields competitive performance even in  the fully parametric situation where the pre- and post-change densities are known to be exponential families.  In that situation, the likelihood-ratio test of \citet{pollak1987average} is expected to be optimal, but we will show that the reduction in power from using our nonparametric test is modest.   Second, we will show that our method outperforms previous methods in the nonparametric case. 

\subsection{Exponential family example}
\label{exp_family_example}

We will first show that, even in an idealized situation where the the parametric form of the pre- and post-change distributions is known, our method can still yield competitive results. We consider the scenario from \cite{pollak1987average}, where $f_0$ is the pdf of a  normal distribution with mean $0$ and standard deviation equal to $6$.  For the post-change density we restrict our attention to shifts in mean. Specifically, we assume that $f_c$  is the pdf of $\text{N}(\mu_c, 6^2)$  where $\mu_c$ is an unknown constant.  At every time,  we observe multiple observations $N_t$.  We use our KS procedure as given in Equation (\ref{continuous_ks}). Thus we work with the raw data $\{y_{t,i}\}$.

\begin{figure}
	\begin{center}
		\includegraphics[width=6in,height= 5.9in]{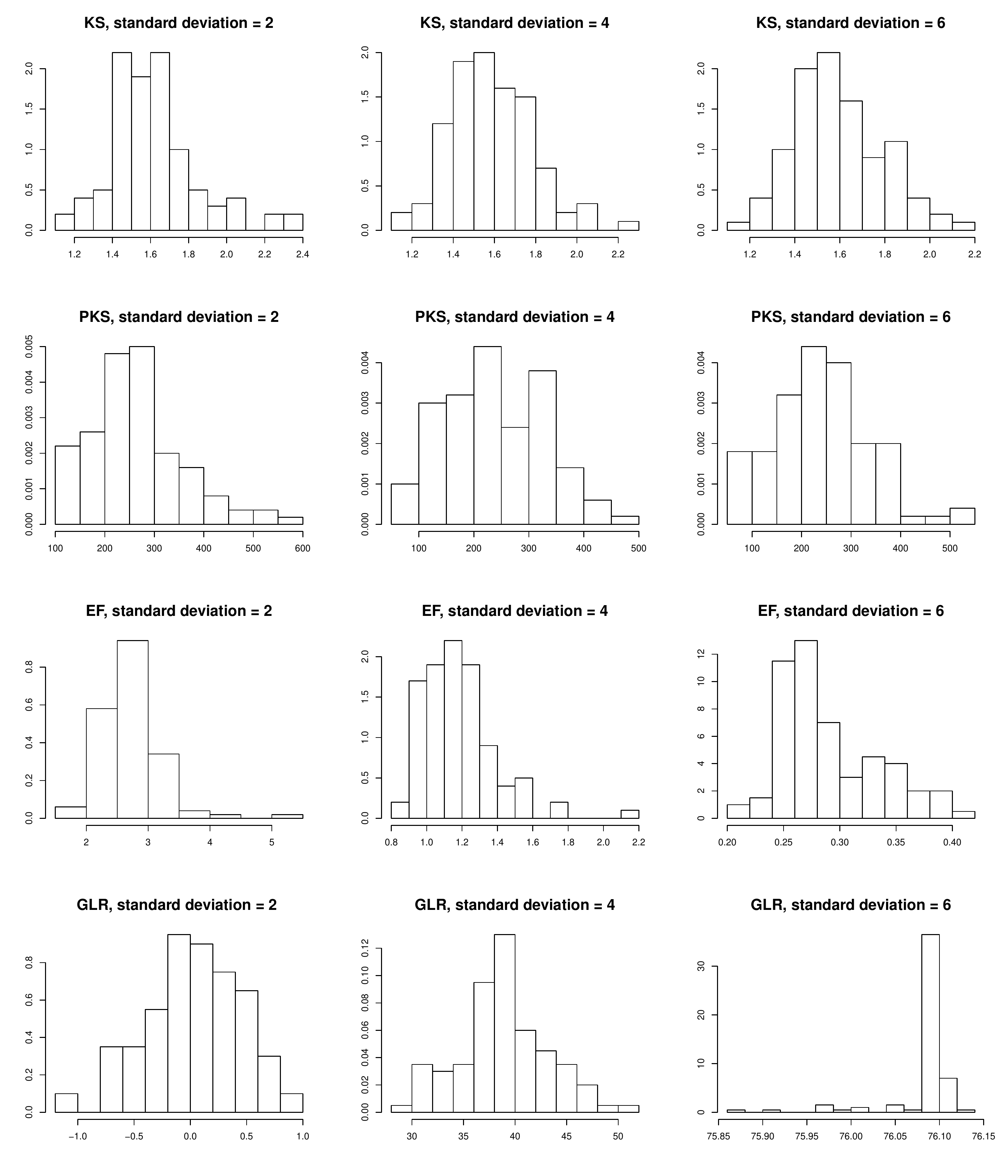} 
		\caption{\label{threshold} Comparisons of the behavior of our KS, predecessor pooled KS, EF, and GLR tests under different null hypotheses.  In each case the null is $N(0, \sigma)$, and the three columns correspond to different values of $\sigma$: 2, 4, and 6, from left to right.  Each panel shows a histogram, derived from 100 Monte Carlo simulations, of the maximum of the test statistic over an interval of length 1000.  Here, EF is computed with $\tau =1$ and under the assumption that after the change point the true mean is unknown and the standard deviation remains unchanged. Panels 10-12 then show the respective comparisons for the method GLR under the assumption that both the mean and standard deviation change after there is a change point.   The KS-based test statistics do not depend on the null (top two rows).  The figure shows that the behavior of the EF and GLR test statistics depends on the null (bottom two rows).  This makes it more difficult to choose a threshold for rejection.}
	\end{center}
\end{figure}

Since the normal distribution has the sample mean as a sufficient statistic for $\mu_c$, we collapse the observations at every time $t$, $\{y_{t,i}\}_{i=1}^{N_t}$,  into their average $\bar{y}_t = N_t^{-1}\sum_{i=1}^{N_t} y_{t,i}$. This simplification allows to use the  exponential family method from \cite{pollak1987average} (EF), and the generalized likelihood-ratio method from \cite{lai2001sequential} (GLR). We recall that these methods were designed under the assumption that at each time instance only one observation arrives. Hence, by working with the averages $\{\bar{y}_t\}$, we can use these methods having as  pre-change density the pdf of $\text{N}(0, 1/N_t)$. Here, for simplicity we assume that $N_t$ is fixed and consider different choices for this value.   In the radiological anomaly-detection problem, the per-period count rate $N_t$ is a proxy for the overall sensitivity of the detector.

\begin{table}
	\centering
	\caption{\label{tab:exponential_family_example}  Average time to detection for different methods under the scenario when both the pre- and post-change densities have a known parametric (in this case Gaussian) form.  EF and GLR refer to the exponential family and generalized-likelihood ratio tests described in the main text.  The rejection thresholds are chosen by Monte Carlo simulation to have an expected value of one false alarm in an interval of length 1000 (except for $\text{KS}^*$ which is our KS  test calibrated as in Section 3.4). We consider different cases of  the post-change mean $\mu_c$.  As expected, the KS-based method does not have as much power to detect false alarms in this situation as methods that exploit the known parametric form of the densities.  }
	\medskip
	\begin{small}
		\setlength{\tabcolsep}{8pt}
		\begin{tabular}{ rrrrrrrrrr }
			\hline
			$\mu_c$ &  $E(N_t)$ &  KS & $\text{KS}^* $  & GLR & $\begin{array}{l}
			\text{EF} \\
			\tau = .1
			\end{array}$  &  $\begin{array}{l}
			\text{EF} \\
			\tau = 1
			\end{array}$  &  $\begin{array}{l}
			\text{EF} \\
			\tau = 5
			\end{array}$ &
			$\begin{array}{l}
			\text{EF} \\
			\tau = 10 
			\end{array}$  \\  
			\hline
			0.1 &  100 & 96.1   & 98.2       & 94.0  &   92.5 & 95.0  & 95.0 & 95.0   \\
			0.1 &  500 & 85.0   & 96.7      & 72.1  &   72.9  &  72.3   & 72.8  & 72.8 \\ 
			0.1 &  1000 &84.4   & 81.3      & 43.2  &   38.7  &39.6   & 39.6  & 39.5 & \\ 			
			0.2 &  100  & 94.5  & 95.7      &  82.2&    80.8  &  80.0& 79.4  &  80.3 \\ 
			0.2 &  500  &  61.6 & 65.0      & 21.9   &   23.6  & 21.0  & 21.1   & 23.7\\ 
			0.2 &  1000  & 32.8 & 34.1      & 12.1 &   12.9  & 12.1   & 12.2  & 13.3\\ 
			0.3 &  100   & 86.3 & 86.8      & 51.3 &   61.0  & 43.2   & 49.9  & 49.9\\ 	
			0.3 &  500    & 33.3& 33.7      & 10.2 &  13.3  & 9.8   & 10.3  & 10.3\\ 
			0.3 &  1000   &  20.4& 23.3     &  5.4 &  7.4  & 5.4   &5.2  & 5.5\\ 
			\hline
		\end{tabular}
	\end{small}
\end{table}

We recall that the EF method requires that we specify a prior for $\mu_c$ under the post-change density, for the computation of the statistic at each time $t$ (which is a Bayes factor).  We assume conjugate normal priors of the form $N(0,\tau^2)$, so that the test statistic can be computed in closed form.  We consider several different choices for the standard deviation under the alternative hypothesis:  $\tau \in \{.1,1,5,10\}$. 

We calibrate methods EF and GLR  by considering a retrospective  window size $L$, similar to the approach taken in our KS procedure and borrowed from \cite{lai1995sequential}. The resulting stopping rules for EF and GLR are given in the appendix.

For our KS method we consider two choices of threshold. The first one corresponds  to calibrate such threshold using Monte Carlo simulations so the expected number of false alarms in an interval of length 1000 equals 1. We denote this procedure by KS. We also consider  a variant, denoted $\text{KS}^*$ which is the one given as Section 3.4 using Corollary \ref{cor1}. Clearly,  $\text{KS}^*$ by construction provides a more conservative rule than $KS$  as the threshold for the  former  comes from an upper bound on the number of false alarms in an interval of a given length and for selected window.

We run comparisons varying the mean shift $\mu_c$. This is shown in Table \ref{tab:exponential_family_example}, where we can see that our KS method  is competitive, but not as efficient  as the EF and GLR methods, which are designed specifically for this case.  It is clear that, if both the pre- and post-change densities belong to a known parametric family, our KS procedure should not be expected to be the best method.   This is consistent with the theory of \cite{pollak1987average}.  As we will show next, however, this is not the case for more general families of distributions.

We do, however, emphasize one practical drawback of the GLR and EF methods, which our method alleviates.  This is shown in Figure \ref{threshold}, where we can see that the distribution of the test statistic under these two methods depends on the null hypothesis (i.e.~it is not a pivotal quantity). This presents a challenge when choosing the threshold for such methods: one must run simulations under the assumed null to choose a threshold for a false alarm.  In contrast, the choice of the threshold for our method does not depend on the null, and can be easily tuned using the theory provided in Section \ref{sec:properties_of_ks}.


\subsection{Mixture of normals}

\label{sec:mixture_normals}

\begin{figure}[t!]
	\begin{center}
		\includegraphics[width=6in,height= 3in]{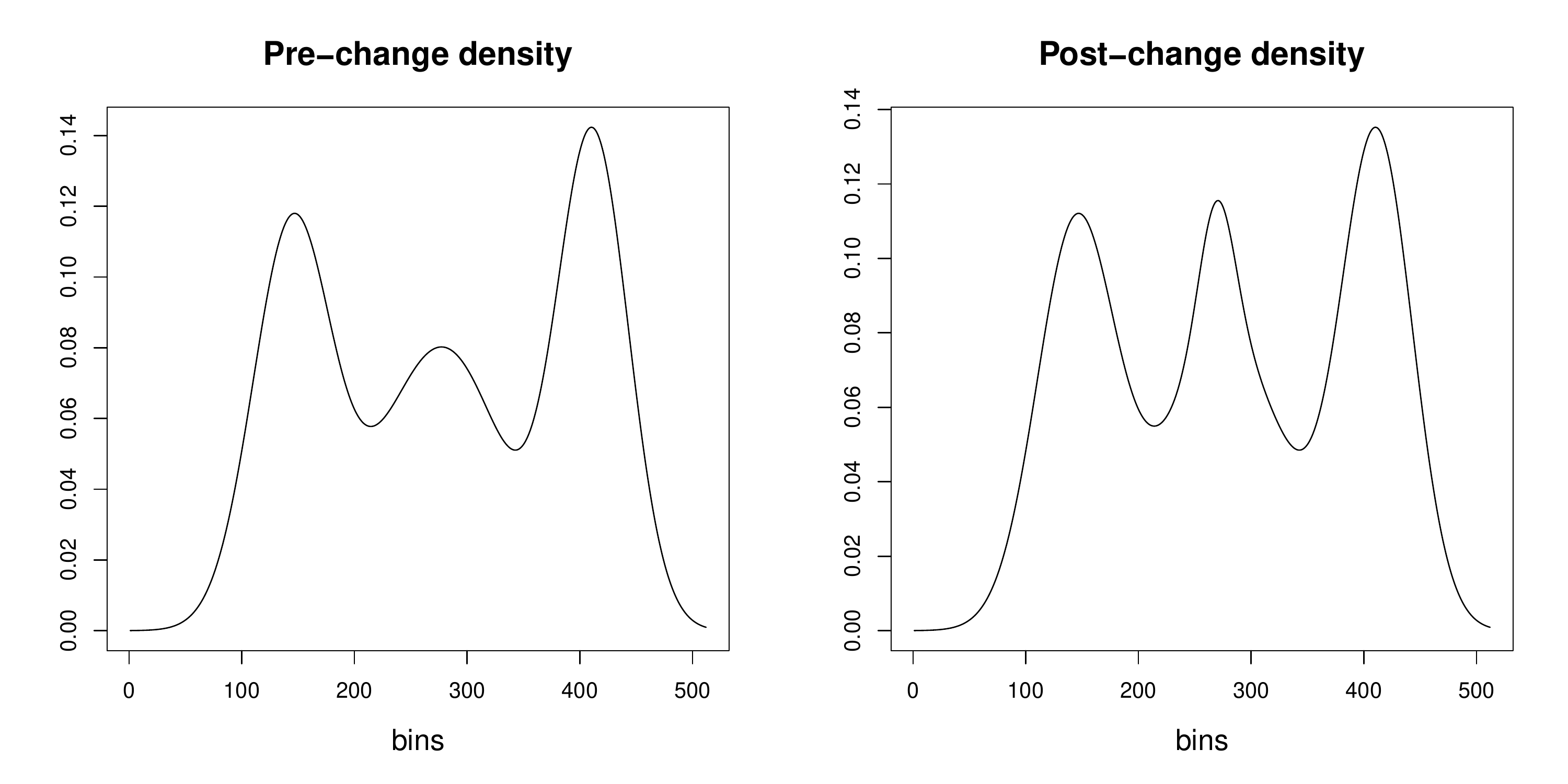} 
		\caption{\label{fig:densities} Pre and post change underlying densities for the mixture-of-normals example in Section \ref{sec:mixture_normals}. }
	\end{center}
\end{figure}

\begin{table}
	\centering
	\caption{\label{tab:mixture_of_normals_example}  Average time to detection for different methods under the mixture-of-normals example. For each scenario, the number of Monte Carlo simulations is 100,  and L=  50.
		For each method the threshold is chosen by Monte Carlo to have an expected value of one false alarm in an interval of length 1000 (except for $\text{KS}^*$ which is our KS  test calibrated as in Section 3.4). We consider different cases of    post change mean $\mu_c$.  PKS is the pooled KS test from \citet{hawkins1988retrospective}. }
	\medskip
		\setlength{\tabcolsep}{18.5pt}
	\begin{small}
		\begin{tabular}{ rrrrrrr }
			\hline
			$D$ &  $E(N_t)$ &  KS & $\text{KS}^*$ & PKS&  EF & GLR \\  
			\hline
			$2^9$ &  100 &  192.6     & 318.8       & 290.3 & 274.8 & 287.8\\
			$2^9$ &  500 &  42.4      & 67.7            & 117.3 & 71.8 & 62.1\\   
			$2^9$ &  1000 & 17.7      & 29.8            & 121.4 &34.5 & 33.2\\ 
			$2^{10}$ &  100 & 152.6   & 195.3            & 197.0 & 317.1 & 257.2 \\
			$2^{10}$ &  500 & 31.2    & 32.2            & 94.7 & 45.2 &  41.0\\   
			$2^{10}$ &  1000  &  12.3 & 15.4            & 64.0 & 31.3 & 30.4 \\ 
			$2^{11}$ &  100 &  60.1   & 94.0            &  130.2 & 289.1 & 192.1\\
			$2^{11}$ &  500 &  12.5   & 16.1            & 81.8 & 35.9 & 36.18\\   
			$2^{11}$ &  1000 & 4.9    & 8.8            & 51.4 & 27.7 & 25.4\\ 			 
			\hline
		\end{tabular}
	\end{small}
\end{table}

We now consider an example where the pre- and post- change densities are not members of a simple parametric family.  We show that in such a situation, our windowed KS test offers a better choice. These findings will be confirmed in the next section with data from field experiments.
 
 To construct the example, we consider a mixture of four normal distributions in the interval $[-8,8]$. Since the methods EF and GLR are not designed  to handle general mixtures of normals, we implement these methods using binned counts of the observations and then making use of the likelihood implied by (\ref{model_before_change}) and (\ref{model_after_change}). The resulting stopping rule for EF is defined as
\begin{equation}
\label{EF_rule}
\inf\left\{t\,\,:\,\,\sum_{s=  \max\{t-L,1\}}^t  \int_{\Lambda} \prod_{k = s}^{t} \frac{p(x_k |  \lambda^{(c)}  )}{p(x_k |  \lambda^{(0)}  )}\,dF(\lambda^{(c)}) \,\geq c  \right\},
\end{equation}
where $dF$ is a the product of independent identically distributed  gamma priors with shape and scale parameters set to $1$.  On the other hand,
the stopping rule for the GLR that we consider here is
\begin{equation}
\label{GLR_rule}
\inf\left\{   t\,\,:\,\,  \max_{ \max\{t-L,1\} \leq k \leq t} \,\,\, \sup_{\lambda^{(c)} \in \Lambda} \,\,\, \sum_{s=k}^t \,\, \log \frac{p(x_s |  \lambda^{(c)}  )}{p(x_s | \lambda^{(0)})} \,\geq c \right\},
\end{equation}
where $\lambda^{(0)}$ is given as in Equation (\ref{model_before_change}).

With this setting, we observe from Table \ref{tab:mixture_of_normals_example}  that the sequential KS method proposed in this paper offers better performance than the other alternatives.  In some cases the increase in power is substantial.  The message of these results is simple: when the nature  of pre- and post- change densities is unknown, then the sequential KS method should be preferred. Even using the more conservative rule from Section 3.4 leads to generally better results than the competing approaches.  With a  real-data example in the next section will further substantiate this claim.


\section{Application to detecting radiological anomalies}

\subsection{Synthetic anomalies derived from real experiments}

\label{sec:data_setup}

It is very challenging to design an operational test of a radiological anomaly-detection system.  Natural experiments are hard to come by, because true radiological anomalies are mercifully rare.  Moreover, for obvious reasons, extensive field tests that involve placing many different radiological anomalies in a crowded urban environment can present an unacceptable risk to public health and safety.

In the next section, we will present the results of one small field experiment conducted in safe, highly controlled circumstances.  But first, we demonstrate the performance of our method at detecting realistic synthetic anomalies.  This study is made as realistic as possible using data collected from two sources:
\begin{enumerate}
\item The 18 hours of background data collected by driving a detector in a golf cart around the University of Texas PRC, as described in the introduction.
\item A small field experiment designed to obtain spectra for two radiological anomalies of the kind that might be encountered in a real law-enforcement setting: cesium-137 and cobalt-60, both of which are in widespread use for industrial and medical applications.  To conduct the experiment, we used small (844 nanoCuries) cesium and cobalt sources provided by the University of Texas's Nuclear Engineering Teaching Laboratory.   We partially shielded the detector from natural background radiation using lead bricks, and we placed the source 5 centimeters away from the detector.   We recorded several minutes of gamma rays from each source, producing detailed spectra.
\end{enumerate}

Thus we have one data source that gives us a background, and another data source that characterizes two likely anomalies.  These give us $f_0$ and two candidates for $f_A$ in Equation \ref{f_c}, respectively.   We use these to construct the synthetic post-change densities in our experiments as follows:
\begin{equation}
\label{eqn:real_post_change}
f_c = \left( \frac{\lambda_0}{ \lambda_0 + \lambda_{\mathrm{source}}  } \right) f_0
+  \left( \frac{\lambda_{\mathrm{source}}}{ \lambda_0 + \lambda_{\mathrm{source}}  } \right) f_A \, ,
\end{equation}
where $\lambda_0$ and $\lambda_{\mathrm{source}}$ are the contributions from the background and the anomaly, respectively.  This is the same setup used in \citet{tansey2015multiscale} when applying the retrospective (i.e.~not sequential) KS test. 

The weights on $f_0$ and $f_A$ are calculated as follows.  Gamma radiation falls off at the rate $1/d^2$ in space, where $d$ is distance to the source, with an additional exponential decay term due to air absorption.  More specifically, as a function of $d$, the contribution of the anomaly to the observed spectrum is
\begin{equation}
\label{eqn:sizetoCPS}
\lambda_{\mathrm{source}} = \frac{\mathrm{mCi}}{0.000844} \cdot 630 \cdot \left( \frac{0.05}{d} \right)^2 \cdot \exp \left\{  -0.0100029(d + 0.05) \right\} \, ,
\end{equation}
where $d$ is distance to the source in meters and $\mathrm{mCi}$ is the size of the source in milliCuries.  Our experiment involving the cesium-137 source was used to calibrate the parameters in Equation \eqref{eqn:sizetoCPS}, based on the measured count rate at 5 centimeters.

\paragraph{Experimental set-up and results.}

\begin{figure}
 		\begin{center}
 			\includegraphics[width=6in,height= 5in]{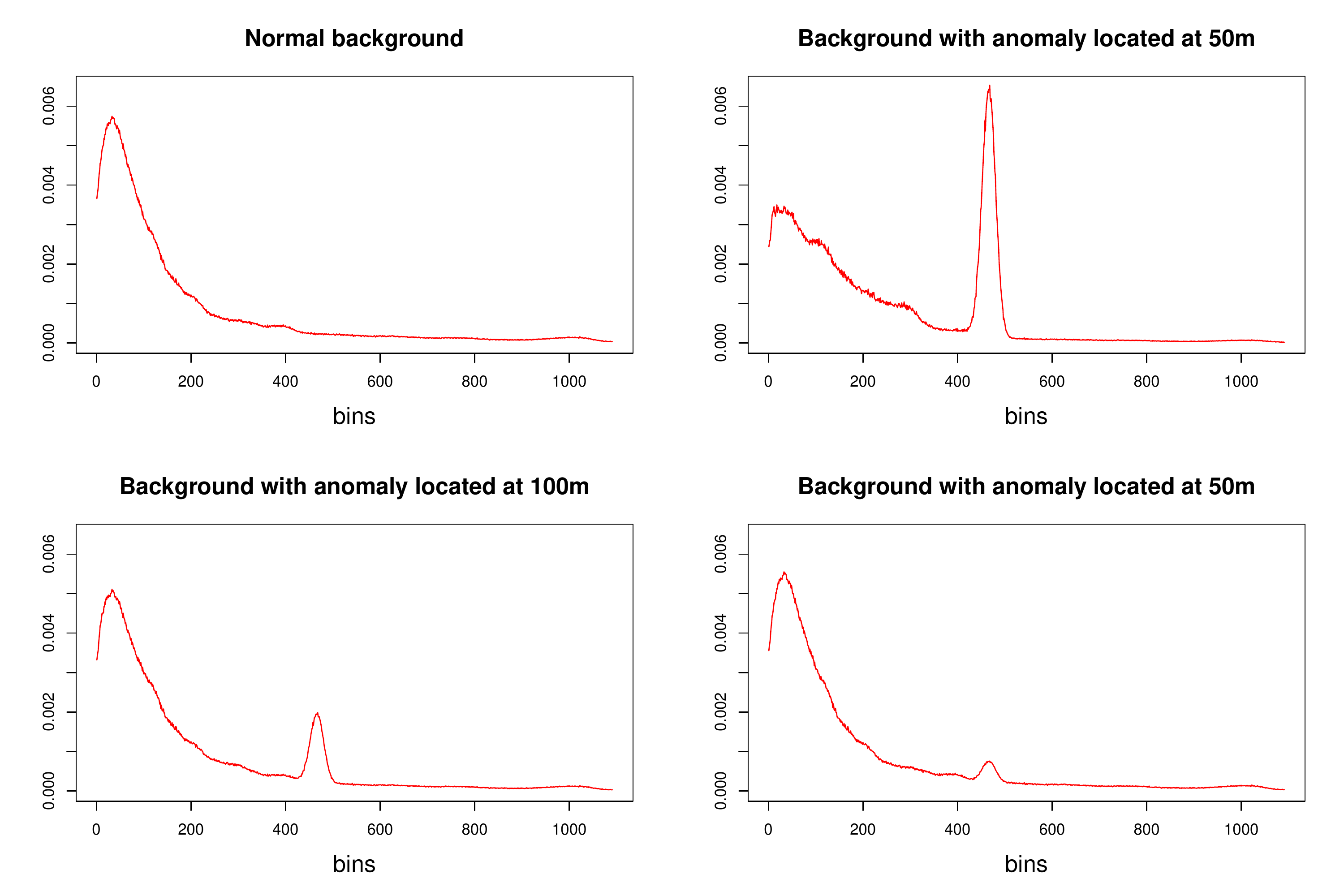} 
 			\caption{\label{fig:celsium} The  first panel above shows the density corresponding to the normal background radiation levels at  PRC (pre--change density). Panels 2-4 then show the spectral distortion of  100 milliCurie Cs-137 source placed at different distances from the radiation detector. }
 		\end{center}
 	\end{figure}

\begin{figure}
 	 		\begin{center}
 	 			\includegraphics[width=6in,height= 5in]{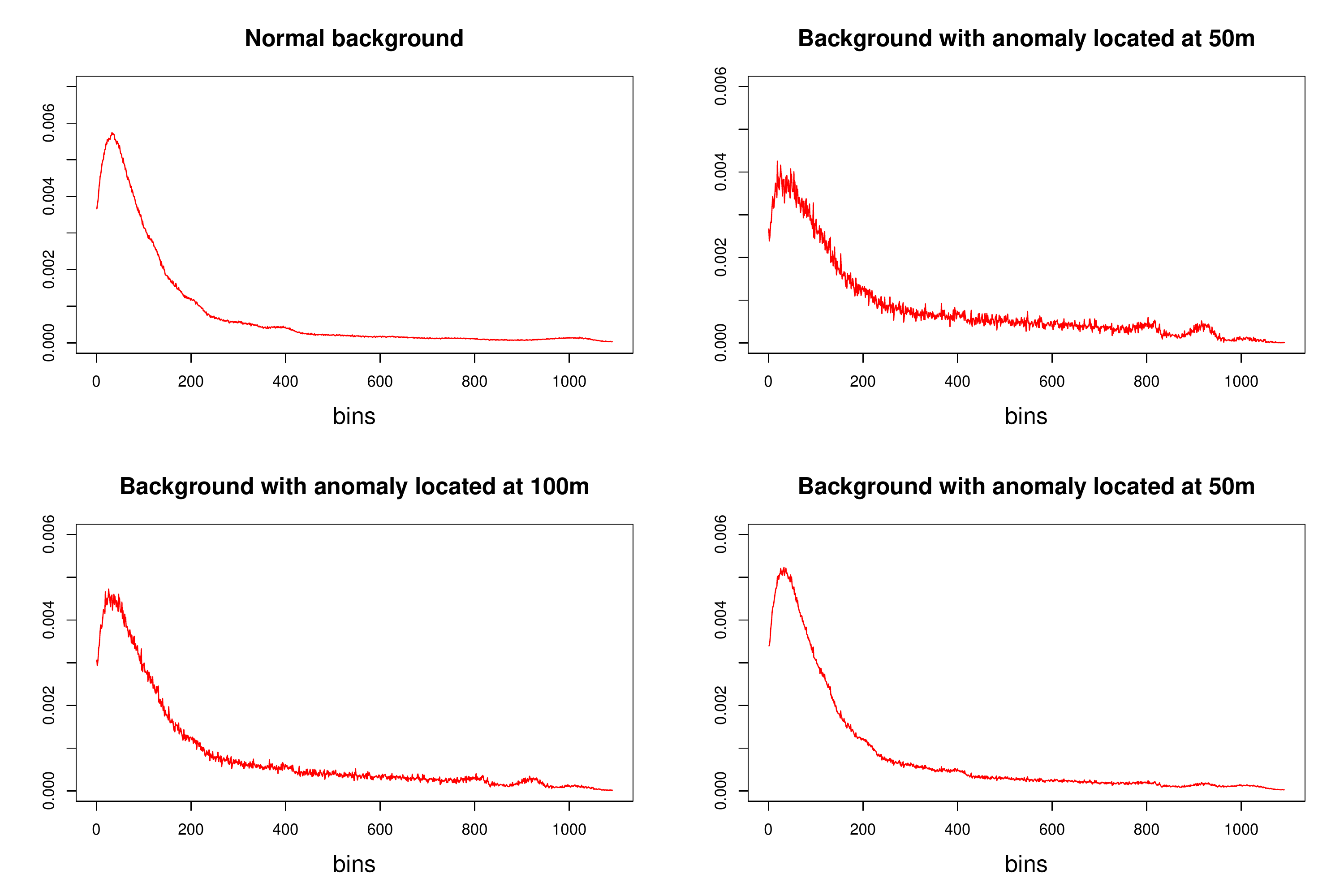} 
 	 			\caption{\label{fig:cobalt}  The  first panel above shows the density corresponding to the normal background radiation levels at  PRC (pre--change density). Panels 2-4 then show the spectral distortion of  650 milliCurie cobalt source placed at different distances from the radiation detector. }
 	 		\end{center}
 	 	\end{figure}
 	
 	\begin{table}[t!]
 		\centering
 		\caption{\label{realisticSimulation} Average time to detection for different methods under the two isotopes considered, Cesium and Cobalt. For each scenario, the number of Monte Carlo simulations is 100,  and L=  50.
 			For each method the threshold is  chosen, using 50 MC simulations, to have an expected value of one false alarm in an interval of length 1000 (except for $\text{KS}^*$ which is our KS  test calibrated as in Section 3.4). We consider three different values for the distance in meters of the observer to the source.}
 		\begin{subtable}{1\textwidth}
 			\setlength{\tabcolsep}{14pt}
 			\small
 			\centering
 			\caption{ Average time to detection for the cesium  examples under the densities in Figure  \ref{fig:celsium}. } 
 			\begin{small}
 				\begin{tabular}{rrrrrrrr}
 					\hline
 					Dist. & $E(N_t)$ & KS & $\text{KS}^*$ & SCR & PKS & EF & GLR  \\ 	
 					\hline
 					50m  & 100     & 1.3    & 1.6     &200       & 19.0  & 7.1   & 5.9  \\
 					50m  & 500     & 1.0    &  1.0    &1.0       & 9.2   & 1.9   & 1.7  \\ 
 					50m  & 1000    & 1.0    &  1.0    &1.0       & 5.9   & 1.2   & 1.0  \\
 					
 					100m & 100     & 9.8    & 12.0     &200       & 66.8  & 24.1  & 19.7 \\ 
 					100m & 500     & 2.6    & 3.1    &8.7       & 30.6  & 9.0   & 8.7   \\    
 					100m & 1000    & 1.6    & 1.8     &1.1       & 19.6  & 6.9   & 6.4  \\
 					  
 					150m & 100     & 111.2  & 161.3     &146.5     & 208.0 & 143.4  & 117.9   \\  
 					150m & 500     & 19.6   & 25.4     &188.7     & 88.8  &28.8    & 27.4 \\ 
 					150m & 1000    & 9.4    & 13.4     &167.3     & 69.7  &18.9    & 18.8 \\ 
 					
 					\hline
 				\end{tabular}
 			\end{small}
 		\end{subtable}\\

		\bigskip
 		\begin{subtable}{1\textwidth}
 			\setlength{\tabcolsep}{14pt}
 			\small
 			\centering
 			\caption{  Average time to detection for the cobalt examples under the densities in Figure  \ref{fig:cobalt}.} 
 			\begin{small}
 				\begin{tabular}{rrrrrrrrr}
 					\hline
 					Dist. & $E(N_t)$ & KS & $\text{KS}^*$  &SCR & PKS & EF & GLR \\
 					\hline
 					50m & 100     & 2.3 &   2.7        &200   & 26.7  & 13.5 & 17.7 \\
 					50m & 500     & 1.0 &   1.0      &21.4  & 12.9  &  7.6 & 8.9 \\ 
 					50m & 1000    & 1.0 &   1.0      &1.0   & 8.1   &  5.1 & 5.2  \\
 					
 					
 					100m & 100    & 5.0 &   5.5       &200   & 44.1   & 21.6   & 28.4 \\  
 					100m & 500    & 1.4 &   1.6       &170.1 & 20.9   & 12.1   & 14.4 \\ 
 					100m & 1000   & 1.0 &   1.1       &7.8   & 13.0   & 10.0   & 9.9   \\  	 
 					
 					
 					150m & 100    & 21.1 &23.9         &200    & 98.6    &  57.0      & 111.5    \\  
 					150m & 500    & 4.9  &5.9         &194.9  & 46.3    &  25.7      & 31.0    \\ 
 					150m & 1000   & 2.9  &3.3         &168.5  & 28.7    &  22.0      & 21.6     \\  	 	 		
 					
 					\hline
 				\end{tabular}
 			\end{small}
 		\end{subtable}\\
 	\end{table}

The post-change densities $f_c$ in our experiments are obtained using Equations (\ref{eqn:real_post_change}) and (\ref{eqn:sizetoCPS}), by varying the source size and the distance to simulate a range of $\lambda_{\mathrm{source}}$ values.  The resulting post-change densities are shown in Figures \ref{fig:celsium} (cesium) and \ref{fig:cobalt} (cobalt).  We benchmark our method against methods EF and GLR described previously, implemented using the Poisson likelihood described in (\ref{model_before_change}) and (\ref{model_after_change}); see Equations (\ref{EF_rule}) and (\ref{GLR_rule}) for the actual stopping rules.  We also benchmark against the pooled KS method (PKS) from Section 3 \citep{hawkins1988retrospective} and the SCR method from \cite{pfund2006examination},

For both the cesium and cobalt anomalies, we considered a range of settings of distance-to-source $d$ and overall detector sensitivity, as parametrized by $E(N_t)$, the expected per-period arrival rate of photons.  We calibrated the rejection threshold of each method so that they all had the same expected false alarm rate of 1 per 1000 time steps. This was done averaging over 100 Monte Carlo simulations, choosing the threshold that produces the desired average false alarms rate. After choosing the threshold, we simulated 100 data sets of length $T=700$.  In each case, the true change point was randomly chosen between $[100,600]$.   All tests with a window size used $L=50$.  To compare the power of each method, we computed the average time to detection: that is, the average number of discrete time steps after the change point until an alarm was raised, across all Monte Carlo simulations.  Lower times to detection indicate higher power.

From Table \ref{realisticSimulation}, we can see that the most effective detection method is our  KS procedure from Section 3.3, which outperforms all other methods across the board.  It offers sometimes dramatic improvements in detection time versus the other methods.  Its superiority is especially marked for the most difficult cases: our method detects the very challenging case of a cobalt-60 source at 150 meters (bottom left panel of Figure \ref{fig:cobalt}) an order of magnitude more rapidly than competing methods.  The SCR method has the most volatile performance.  It performs about as well as our KS test at detecting easy anomalies like the upper right panel of Figure \ref{fig:celsium}.  But it performs much more poorly for anomalies that are difficult to detect.

\begin{figure}[t!]
	\begin{center}
		\includegraphics[width=6in,height= 3.5in]{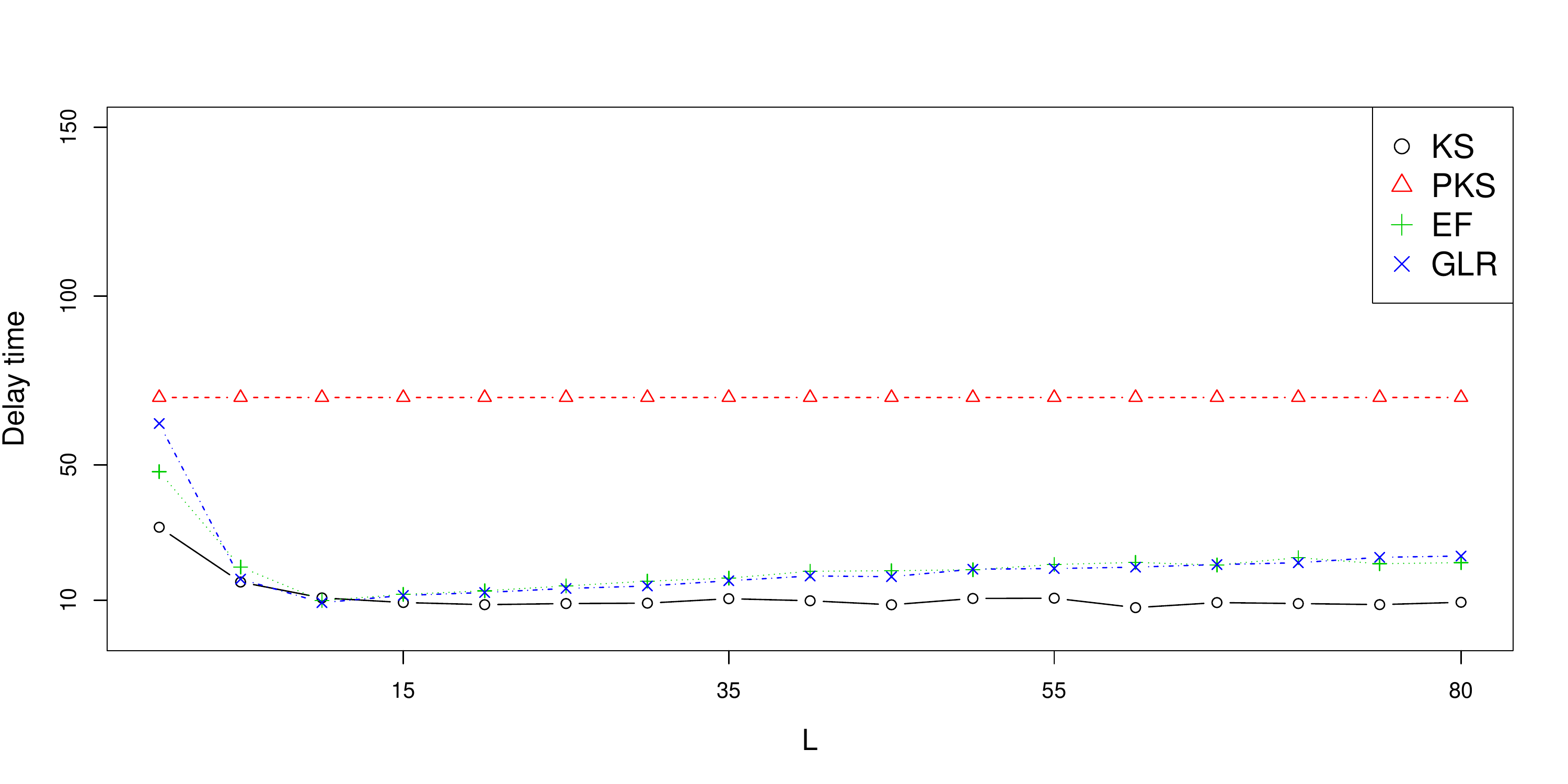} 
		\caption{\label{vary_L} Average time to detection comparisons, over 100 MC simulations, when the parameter $L$ varies. Here  the pre- and post- change densities are given as in Figure \ref{fig:celsium} where post--change density corresponds to an anomaly located at a distance of 100 m, and setting the intensity of photon measurements as $E(N_t) = 100$. Moreover,  for each method the	threshold is  chosen, using 100 MC simulations, to have an expected value of one false alarm in an interval of length 1000. Note that PKS behaves like constant as it does not require to choose $L$. It is plotted with the purpose of a more complete comparison. 
			  }
	\end{center}
\end{figure}

As a robustness check, we assess the sensitivity of different methods to the choice of   window size $L$. This is depicted in Figure \ref{vary_L}, where it is clear that our KS sequential procedure is the most efficient. Here we see that, although larger values of $L$ typically produce better results, the gain becomes smaller as $L$ increases. From our experience with these experiments, and other informal experiments not described here, we have found that $L \approx 50$ is a reasonable choice for the situations considered in this paper. 

\subsection{A real anomaly}

\begin{figure}
	\begin{center}
		\includegraphics[width=6in]{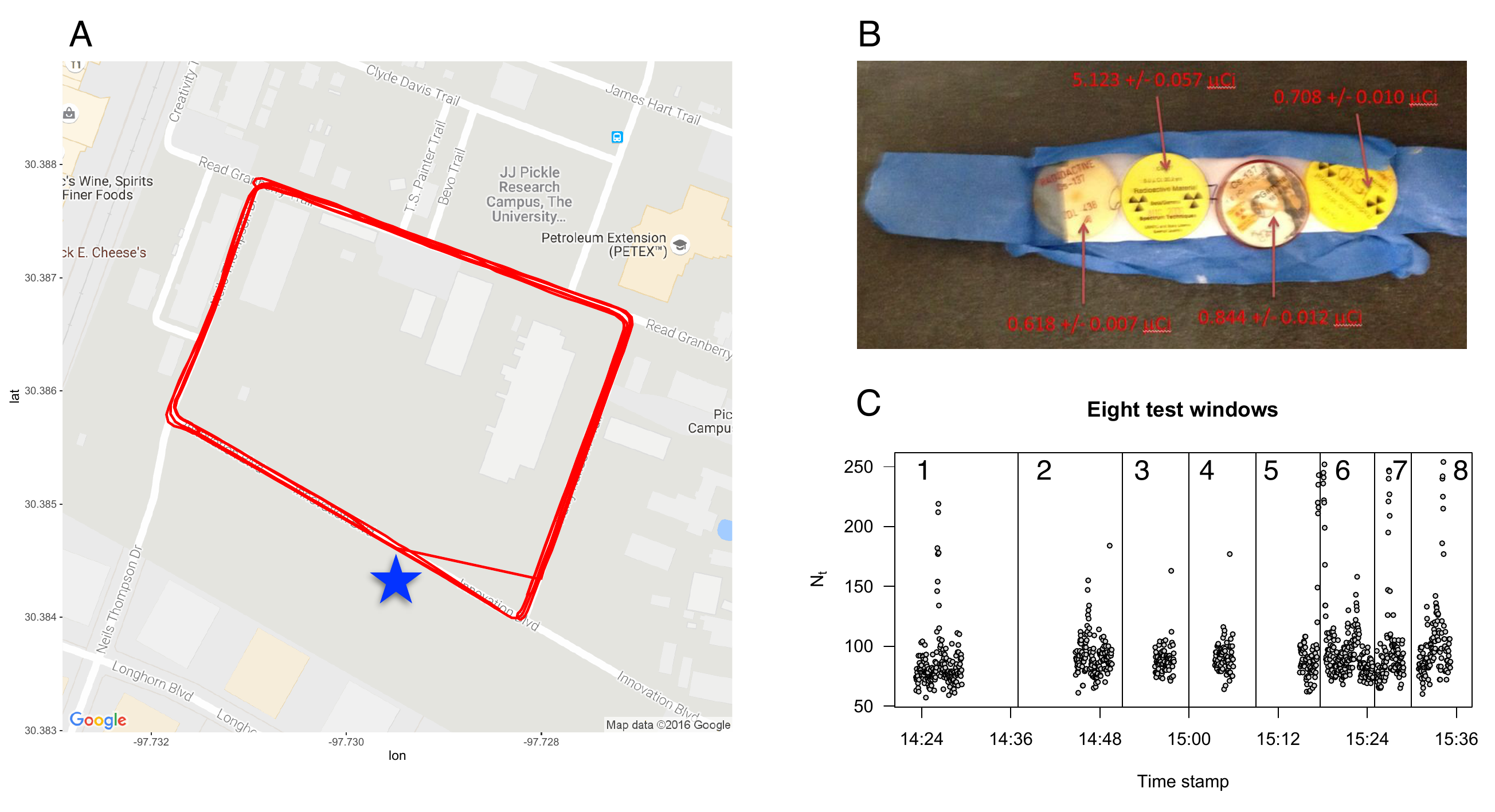}
		\caption{\label{time_stamp} The setup for the real anomaly-detection experiment, in which a radiation detector was driven around in a golf cart in an area where a known (small, harmless) radiological anomaly had been placed.  Panel A shows a map of the test site, the GPS trace of the golf cart's path (red lines), and the approximate location of the anomaly (blue star).  Panel B shows a photograph of the four small laboratory check sources bundled together to form the anomaly.  Panel C shows the eight different test windows created from the 33 minutes of recording.  Windows 1, 2 and 6 each involve one full rectangular circuit of the roads around the anomaly.  Windows 3, 4, 5, 7 and 8 each involve a half circuit.   The $y$ axis in Panel C is $N_t$, the number of gamma rays detected in each discrete 2-second time step.}
	\end{center}
\end{figure}

We conclude with an analysis of a small, controlled field experiment.  A small, harmless (6.293 micro-Curie) check source of radioactive cesium-137 was obtained from University of Texas's Nuclear Engineering Teaching Laboratory and placed at a known location on the University of Texas Pickle Research Campus, twelve inches above the ground.  (See Panels A and B of Figure \ref{time_stamp}.)  A radiation detector that outputs gamma-ray measurements in discrete 2-second intervals was then placed in a golf cart and driven several times around an approximately rectangular circuit of the nearby roads, passing within ten feet of the source at speeds under 10 mph.  This protocol yielded approximately 33 minutes of data collected in the presence of an anomaly (see Figure \ref{time_stamp}).  The background spectrum $f_0$ at this location had been very well characterized from the month-long data-collection effort described in the introduction, and could be treated as known.

We separated the 33 minutes of data into eight different independent test windows of varying lengths: three large-length windows comprising a full circuit of the testing site, and five shorter windows comprising a half circuit.  Here a ``circuit'' means one full lap of the golf cart around the testing site, as shown in Panel A of Figure \ref{time_stamp}.  We treat each window shown in Panel C of Figure \ref{time_stamp} as a separate test set.

To design our  comparisons, we use historical data to estimate $f_0$. We also use different historical data  as held out sets  to calibrate the 
the false alarm threshold for the methods EF, PKS, and GLR, and  KS test. However, here we only use held out sets that are not rejected when testing a one time KS test under the null $f_0$. This is done since the data historical data was collected over different locations, and days of the year. Moreover, we consider  $\text{KS}^*$ which is our KS test calibrated as in Section 3.4.  To reflect the relatively small size of the data set, we set the expected rate of false alarms is set to be 1 in an interval of 175 time steps.  We refer to test windows shown in Figure \ref{time_stamp}  as 1--8 according to the  order in which they were collected. 

 \begin{table}
 	\centering
 	\caption{\label{tab:real_example} Time to detection (measured by the number of discrete two-second time steps required to raise an alarm for different methods in our real-data experiment.  We denote the detection time by $\infty$ for any method which was not able to detect the anomaly in a given instance.  }
 	\medskip
 	\setlength{\tabcolsep}{18.5pt}
 	\begin{small}
 		\begin{tabular}{ rrrrrrrr }
 			\toprule
			$\begin{array}{l}
				\text{Test}\\
				\text{ window}
			\end{array}$ &  KS &  $\text{KS}^*$  &PKS&  EF & GLR  &  SCR\\  
 			\midrule  
			1 &  16 &  16   &16     & 24       & 103         &88\\ 
			2 &  8  &  19   &19     & 22       & 22          & 124\\
			3 &  9  &  9    &23     & 56       & 56          &$\infty$\\   
			4 &  17 &  17   &25     & $\infty$ &$\infty $     &$\infty$\\ 			 
			5 &  55 &  55   &43     & 76       &76              &76\\ 
			6 &  5  &  5    &6      & 12       &12              &147\\
			7 &  16 &  17   &16     & 52       &52              &49\\ 
			8 &  29 &  29   &22     & 98       &97              &95\\ 
 			\bottomrule
 		\end{tabular}
 	\end{small}
 \end{table}

Table \ref{tab:real_example} shows the results of the experiment.  Overall, our KS procedure is the best-performing method, confirming what we observed in our previous experiments.  Our method detects the anomaly in all eight test windows, and is either the fastest or second-fastest to do so in each case.  The PKS method also detects the anomaly in all eight windows, but for most cases not as quickly as our method does.  The EF and GLR methods detect the anomaly in seven out of eight cases, while the SCR method fails to detect the anomaly in two cases.

\section{Discussion}

We have proposed a sequential nonparametric test for a change in distribution and have characterized both its theoretical properties and its practical performance versus plausible alternatives.   The method can be deployed easily in any streaming-data scenario, where retrospective batch tests do not match the design requirements of the anomaly-detection system.  Our theoretical results, as well as our experiments on both real and simulated data, demonstrate that the method is both robust and immediately useful for the problem of radiological anomaly detection.

While we have emphasized the mobile-detector scenario in this paper, our method could also be applied in the fixed-detector scenario (e.g.~at ports or border crossings).  A complication here is that stationary detector systems, while they could benefit from our method as well, would involve extra methodological features to ignore anomalies caused by known benign radioactive sources, like truckloads of rock salt or granite \citep[e.g.][]{runkle:etal:2009}.  Combining such a system with our approach is an active area of research.

One final point worth noting is that real radiological data can have nuisance components, and that the counts collected with an spectrometer could violate the Poisson-distribution assumption. In this case, Theorem \ref{theorem1} and Corollary \ref{cor1} would not hold as stated here, but would have to be adjusted to account for overdispersion (this was not the case with our real data).  In situations where the over-dispersion is constant, as assumed by \cite{chan2014distribution}, our statistic can easily be generalized, with no additional complication. Moreover, the over-dispersion parameter can be  estimated consistently, using background-only data, as in \cite{chan2014distribution}.



\paragraph{Acknowledgments.} The authors thank Steven Biegalski of the Nuclear
Engineering Teaching Laboratory for assistance in obtaining the cesium-137
source and in performing the controlled experiment, and for valuable advice.

\begin{small}
	\singlespacing
	\bibliographystyle{abbrvnat}
	\bibliography{anomaly_detection}
	
\end{small}


\appendix

\section{Proof of technical results}

\subsection*{Proof of Theorem \ref{theorem1}}

\begin{proof}
	First, we note that by the triangle inequality
        \begin{equation}
        \begin{aligned}
	\label{eq1_t1}
        \left \vert \frac{\Delta_{s,t}}{\sqrt{ \sum_{k=s}^{t} N_k }}  - d(F_c,F_0)\,  \right\vert  & = \left\vert  \max_{1 \leq j \leq D} \vert F^0(j) - \hat{F}^{s:t}(j) \vert  \,\,-\,\,  \max_{1 \leq j \leq D} \vert F^0(j) - F_c(\xi_j) \vert    \right\vert\\
        & \leq   \max_{1 \leq j \leq D}\,  \left\vert \vert F^0(j) - \hat{F}^{s:t}(j) \vert -  \vert F^0(j) - F_c(\xi_j) \vert  \right\vert\\
        & \leq \max_{1 \leq j \leq D}\, \vert F_c(\xi_j) - \hat{F}^{s:t}(j) \vert \\
        & \leq  \sup_{\xi \in \mathbb{R}}\, \left\vert F_c(\xi) - \hat{F}_{s:t}(\xi)  \right\vert.
      \end{aligned}
    \end{equation}
	Next we observe that since $\sum_{j=1}^{D} \lambda_j^c > 0$ and $s> v $ then
	
	\[
	P\left(  \sum_{k=s}^{\infty}N_k  \,=\, \infty \right) \,=\, P\left(  \sum_{k=s}^{\infty}\sum_{j =1}^{D} x_{k,j}  \,=\, \infty \right) \,=\,1
	\]
	hence by the Glivenko-Cantelli theorem we arrive to
	
	\[
	P\left(\lim_{t \rightarrow \infty}  \frac{\Delta_{s,t}}{\sqrt{ \sum_{k=s}^{t} N_k }}  \,= \, d(F_c,F_0)\right) = 1,
	\]
	and the claim follows since $ d(F_c,F_0) > 0$.
	
	On the other hand, from (\ref{eq1_t1}), if
	
	\[
        D_{s,t}^{\prime} := \sqrt{ \sum_{k=s}^{t} N_k }\,\,\sup_{\xi \in \mathbb{R}}\, \vert F_c(\xi) - \hat{F}_{s:t}(\xi) \vert,
	\]
	by Corollary 1 from  \cite{massart1990tight} we obtain
        \begin{align*}
        \text{P}\left(\left\vert \Delta_{s,t}  - d(F_c,F_0)\,\sqrt{
          \sum_{k=s}^{t} N_k }  \right\vert   \leq c_L \right)   & \geq  \text{P}\left(  D_{s,t}^{\prime}  \leq c_L
	\right) \\
         & =  \sum_{n_s= 1}^{\infty} \ldots \sum_{n_t = 1}^{\infty} \text{P}\left( D_{s,t}^{\prime}  \leq c_L
	, N_s = n_s,\ldots,N_t = n_t\right) \\
        & \geq (1- 2\,\exp(-2\,c_L^2)) \sum_{n_s,\ldots, n_t}  \text{P}\left(N_s = s,\ldots,N_t = n_t\right)  \\
        & = (1- 2\,\exp(-2\,c_L^2)).
        \end{align*}
        and the result follows.
	%
	%
	

\end{proof}

\subsection*{Proof of Corollary  \ref{cor1}  }

For the first part simply note that

\begin{align*}
E\left(A_T \right)  &=  E\left(  \sum_{t=1}^T \boldsymbol{1}_{ \left\{   \max_{  \max\{t-L,1\} \leq s \leq t } \Delta_{s,t}   \geq  c_L  \right\} }   \right) \\
&\leq  \sum_{t=1}^T E\left(  \sum_{s =\max\{t-L,1\} }^t \boldsymbol{1}_{ \left\{ \Delta_{s,t}   \geq   c_L    \right\} }   \right)  \\
& = \sum_{t=1}^T\,\sum_{s =\max\{t-L,1\} }^t P\left( \Delta_{s,t} \geq  c_L   \right)   \\
& = \sum_{t=1}^T\,\sum_{s =\max\{t-L,1\} }^t P\left( D_{s,t} \geq  c_L   \right)   \\
& \leq 2\,T\,L\,\exp\left( -2\, c_L  ^2  \right)
\end{align*}
where the last inequality follows from  Corollary 1 in  \cite{massart1990tight}.

For the second part we observe that

\begin{align*}
P\left(  \tau_{L} \leq t \mid \tau_{L}  > v \right)  \geq{} & P\left(   \max_{\max\{t-L,1\} \leq s \leq t  } \Delta_{s,t} \geq c_L \mid \tau_{L}  > v \right) \\
 \geq{} &  P\left(   \Delta_{v+1,t} \geq c_L \mid \tau_{L}  > v \right)  \\
={} &  P\left(   \Delta_{v+1,t} \geq c_L \right)  \\
\geq{} &  P\left(   \Delta_{v+1,t} \geq -c_L   + d(F_c,F_0)\,\sqrt{ \sum_{k=v+1} ^t N_k } ,  d(F_c,F_0)\,\sqrt{ \sum_{k=v+1} ^t N_k }  \geq  2\,c_L\right) \\
\geq{} &  1- P\left(  \Delta_{v+1,t}  <  -c_L + d(F_c,F_0)\,\sqrt{ \sum_{k=v+1} ^t N_k } \right)    \\
& {}- P\left( \,\sqrt{ \sum_{k=v+1} ^t N_k }  < \frac{2\,c_L}{d(F_1,F_0)}  \right)\\
\geq{} & 1 -  P\left(   c_L < \left\vert  d(F_c,F_0)\,\sqrt{ \sum_{k=v+1} ^t N_k } - \Delta_{v+1,t} \right\vert \right) \\
& {}- P\left( \,\sqrt{ \sum_{k=v+1} ^t N_k }  < \frac{2\,c_L}{d(F_1,F_0)}  \right)\\
={} &   P\left(   \left\vert  d(F_c,F_0)\,\sqrt{ \sum_{k=v+1} ^t N_k } - \Delta_{v+1,t} \right\vert  \leq c_L \right) \\
&  {}- P\left( \,\sqrt{ \sum_{k=v+1} ^t N_k }  < \frac{2\,c_L}{d(F_1,F_0)}  \right)
\end{align*}
and the claim follows once more by  Corollary 1 in  \cite{massart1990tight} as in the previous Theorem.

\subsection{EF  and GLR  stopping rules for Section \ref{exp_family_example} }

The stopping rule for the EF method is given by 
\begin{equation}
\label{EF_rule2}
\inf\left\{t\,\,:\,\,\sum_{s=  \max\{t-L,1\}}^t  \int_{\mathbb{R}} \prod_{k = s}^{t} \frac{N(\bar{y}_k |  \mu^{(c)},(6/\sqrt{N_t})^2  )}{N(\bar{y}_k |  0, (6/\sqrt{N_t})^2  )}\,N(\mu^{(c)} | 0,\tau^2)\,d\mu^{(c)} \,\geq c  \right\},
\end{equation}
where $dF$ is a the product of independent identically distributed  gamma priors with shape and scale parameters set to $1$.  On the other hand,
the stopping rule for the GLR that we consider there is
\begin{equation}
\label{GLR_rule2}
\inf\left\{   t\,\,:\,\,  \max_{ \max\{t-L,1\} \leq k \leq t} \,\,\, \sup_{\mu^{(c)} \in \mathbb{R}} \,\,\, \sum_{s=k}^t \,\, \log \frac{N(\bar{y}_s |  \mu^{(c)},(6/\sqrt{N_t})^2  )}{N(\bar{y}_s | 0,(6/\sqrt{N_t})^2)} \,\geq c \right\}.
\end{equation}


\end{document}